\documentclass[12pt]{article}
\usepackage{amsmath}
\usepackage{graphicx,psfrag,epsf}
\usepackage{enumerate}
\usepackage[round]{natbib}
\usepackage{url} 
\usepackage{amsmath,amssymb,amsfonts,amsthm}
\usepackage{float,graphicx, multirow, setspace}
\usepackage{subfiles}
\usepackage{placeins}
\usepackage{url}
\usepackage{hyperref}
\usepackage{xr, xr-hyper}
\usepackage{algpseudocode}
\usepackage{subfig}
\usepackage{algorithm}
\usepackage{makecell}
\usepackage[scr=boondox]{mathalfa}

\usepackage{afterpage}

 \newcommand{\bE}{{\mathbb E}}
 \renewcommand{\P}{{\mathbb P}}
 \newcommand{\ind}[1]{\mathbb{I}\left(#1\right)}

 \newcommand{\bx}{{ \mathbf{x}}}
 \newcommand{\bX}{{ \mathbf{X}}}
 \newcommand{\bXtrain}{{ \mathbf{X}_{\text{train}} }}
 \newcommand{\bw}{{ \mathbf{w}}}
 \newcommand{\by}{{ \mathbf{y}}}
 \newcommand{\bytrain}{{ \mathbf{y}_{\text{train}} }}
 \newcommand{\bwtrain}{{ \mathbf{w}_{\text{train}} }}
  \newcommand{\D}{{\mathscr{D} }}
  \newcommand{\Dtrain}{{\mathscr{D}_{\text{train}}}}
  \newcommand{\Dtest}{{\mathscr{D}_{\text{test}}}}
\newcommand{\Ndrives}{  N_{\text{drives}} }
\newcommand{\Mtest}{  M_{\text{test}} }

 \newcommand{\SD}{{ \mathsf{SD}}}
 \newcommand{\SE}{{ \mathsf{SE}}}
 \newcommand{\rmse}{{ \mathsf{rmse}}}
 \newcommand{\logloss}{{ \mathsf{logloss}}}
 \newcommand{\covg}{{ \mathsf{covg}}}
 \newcommand{\bootcovg}{{ \mathsf{bootcovg}}}
 \newcommand{\predset}{{ \mathsf{predset}}}
 \newcommand{\predsetp}{{ \predset_{\bphat} }}
 \newcommand{\predsetpboot}{{ \predset_{\bphat^{(1)},...,\bphat^{(B)}} }}

\newcommand{\xgb}{{ \mathsf{XGBoost}}}

 \newcommand{\pts}{{ \mathsf{pts}}}
 \newcommand{\ep}{{ \mathsf{EP}}}
 \newcommand{\phat}{{ \widehat{p} }}
 \newcommand{\bphat}{{ \widehat{\mathbf{p}} }}
 \newcommand{\bphatCat}{{ \widehat{\mathbf{p}}_{\text{prior}} }}
 \newcommand{\bphatTarget}{{ \widehat{\mathbf{p}}_{\text{target}} }}
 
 \newcommand{\ephat}{{ \widehat{\ep} }}
 \newcommand{\epa}{{ \mathsf{EPA}}}

 \newcommand{\R}{{\mathsf R}}
 \newcommand{\nflfastr}{{\textsf{nflFastR}}}

 \DeclareMathOperator*{\argmin}{arg\,min}
\usepackage[dvipsnames]{xcolor}

\newcommand{\blind}{0}

\begin{document}

\bibliographystyle{apalike}

\def\spacingset#1{\renewcommand{\baselinestretch}%
{#1}\small\normalsize} \spacingset{1}


\if0\blind
{
  \title{
    \bf{Moving from Machine Learning to Statistics: the case of Expected Points in American football}
  }

  \author{
    Ryan S. Brill\thanks{
      Graduate Group in Applied Mathematics and Computational Science, University of Pennsylvania. Correspondence to: ryguy123@sas.upenn.edu
    }, \ 
    Ryan Yee\thanks{
      Dept.~of Statistics, University of Wisconsin--Madison
    }, \ 
    Sameer K. Deshpande\footnotemark[2], \ 
    and Abraham J. Wyner\thanks{
      Dept.~of Statistics and Data Science, The Wharton School, University of Pennsylvania
    }
  }
  \maketitle
} \fi

\if1\blind
{
  \bigskip
  \bigskip
  \bigskip
  \begin{center}
  \title{
    \bf{A statistical view of expected points models in American football}
  }
  \end{center}
  \medskip
} \fi

\bigskip
\begin{abstract}
Expected points is a value function fundamental to player evaluation and strategic in-game decision-making across sports analytics, particularly in American football. 
To estimate expected points, football analysts use machine learning tools, which are not equipped to handle certain challenges. 
They suffer from selection bias, display counter-intuitive artifacts of overfitting, do not quantify uncertainty in point estimates, and do not account for the strong dependence structure of observational football data.
These issues are not unique to American football or even sports analytics; they are general problems analysts encounter across various statistical applications, particularly when using machine learning in lieu of traditional statistical models.
We explore these issues in detail and devise expected points models that account for them. 
We also introduce a widely applicable novel methodological approach to mitigate overfitting, using a catalytic prior to smooth our machine learning models.


\end{abstract}

\noindent%
\vfill

\newpage
\spacingset{1.45} 


\section{Introduction}\label{sec:intro}

Sports analytics has become a multibillion-dollar industry.
Each team in Major League Baseball (MLB), the National Basketball Association (NBA), and the National Football League (NFL) has at least one analytics staffer.
Many of these teams have full-fledged sports analytics research groups
and hire sports analytics consulting firms.
The popularity of these firms, many of which consist of large teams of Ph.D.s in statistics, mathematics, and computer science, has exploded recently.

Two fundamental areas of interest to these teams and firms are player evaluation and in-game strategic decision-making.
The quantitative approach relies on a valuation function, typically an expected value, that measures the value of each game-state. 
Analysts can evaluate an individual player by the value added across each of his plays.
They can also evaluate a coach by how often he makes decisions that maximize the value of the next game-state.

The most prominent and widely used value function across all of sports analytics is expected points ($\ep$).
In baseball, $\ep$ (or expected runs) is the expected number of runs scored from the current game-state through the end of the half-inning.\footnote{
  \url{https://library.fangraphs.com/misc/re24/}
}
In American football, which runs in continuous time, $\ep$ is the expected net number of points scored from the current game-state through the next scoring event (or the end of the half) \citep{nflWar}.
In soccer, $\ep$ (or expected possession value) is the likelihood that the team with possession of the ball scores the next goal minus the likelihood it concedes the next goal, given the current game-state \citep{Fernandez}.
$\ep$ is defined similarly for any sport with scoring and time components.

In this work, we focus on $\ep$ for American football as our primary case study for several reasons. 
First, the creation and development of expected points methodologies in football has been and continues to be an open source endeavor \citep{VirgilCarter71,carroll1989hidden,Romer06dofirms,Burke4thDownPt1}.
State of the art $\ep$ models today are open source and reproducible from publicly available data \citep{nflWar,BaldwinWP,nflFastR}.
In contrast, soccer $\ep$ methodologies and data aren't fully publicly available.
All state of the art analyses are proprietary -- fit by teams or by collaborations with consulting firms\footnote{
  e.g., \url{https://statsbomb.com/soccer-metrics/possession-value-models-explained/}
} from proprietary tracking data.\footnote{
  e.g., \url{https://www.statsperform.com/team-performance/football-performance/sportvu/}
} 
Second, American football analysts often use expected points to evaluate players in the public sphere, e.g. by posting player or team rankings online.
Analysts commonly rank team offenses, team defenses, and quarterbacks across a season using the expected points added metric ($\epa$), which is computed by subtracting the  expected points at the beginning of a play from the expected points at the end of a play \citep{nflWar}.
Player or team performance is often measured with respect  to an average performance baseline, which is accomplished by  assuming that an offense faces an average defense and that a defense faces an average offense.

The formulation of expected points for American football begins with a partitioning of a game into groups of approximately independent segments we call \textit{epochs}.
An epoch ends with a score, the end of the first half or the end of the game. 
Each play within an epoch is characterized by a detailed game-state vector $\bx$ (e.g., third down and eight yards to go from the opponent's 15 yard-line with two minutes left in the first half, etc.).
We are interested in estimating  $\ep(\bx)$, the expected number of net points scored at the end of the epoch relative to the team that currently possesses the ball.
Expected points is not a counting statistic\footnote{
  A counting statistic can easily be calculated by a viewer by counting the number of specific events or actions that occur over a period of time.
  Examples from baseball include hits, home runs, runs, total bases, strikeouts, and walks.
  Anyone watching a game can determine counting statistics simply by tabulating these box score events.
} readily observable or calculable by a viewer -- it requires and is defined entirely with respect to a model. 
It is a football analyst's challenge to estimate the expected net points scored at the end of an epoch.

In principle, one could estimate $\ep(\bx)$ by finding all instances of state $\bx$ and taking the average net points of the next score.
Such binning and averaging has a long history and has been widely used.
In baseball, this technique underlies run expectancy with 24 game-states (RE24).\footnote{
  \url{https://library.fangraphs.com/misc/re24/}
} 
Binning and averaging works well in baseball because there are a small number of influential game-state variables (i.e., base runners and outs) that impact expected runs.
Also, there are many independent observations for each state.
This is not true in American football, so binning and averaging is inaccurate and impractical.
As we account for more variables, the number of game-states increases multiplicatively. 
In football, the number of possible game-states is prohibitively large.
For instance, all combinations of just yardline, down, and seconds remaining in the half account for over 7 million possible game-states.
Further, football has many fewer plays per season than baseball.
Thus, the sample size of each game-state is orders of magnitude smaller in football than in baseball.

To overcome these challenges, analysts initially formulated $\ep$ as a regression problem estimated at the level of plays.
That is, analysts fit a statistical model that imposed a simplified structure on the valuation function in order to borrow strength across similar game-states.
Early approaches from \citet{Romer06dofirms} and \citet{Burke4thDownPt1} used simple models of the net points of the next score as a function of game-state.
The former used instrumental variable regression and the latter used spline regression.
\citet{nflWar} later modeled the probability of each potential outcome of the next score via multinomial logistic regression and accounted for many more game-state variables.
They then estimated expected points as a weighted sum of these outcome probabilities.

Recently, blackbox machine learning algorithms have become the dominant way to fit complex functions across sports analytics, as they have in countless other applications.
Their widespread popularity and adoption was catalyzed by their modeling flexibility, the accessbility of rich, pubicly available data (e.g., play-by-play data from nflFastR \citep{nflFastR}), and off-the-shelf machine learning tools (e.g., $\xgb$ \citep{xgboost}). 
These models are widely assumed to be reasonable and trustworthy.
Accordingly, football analysts have recently turned to machine learning to estimate expected points.
Notably, \citet{nflFastREPxgb} fit a widely used expected points model using $\xgb$.

Modern machine learning approaches, however, overlook several extremely important and quite thorny statistical issues.
They don't adjust for team quality, which produces $\ep$ estimates that are then misinterpreted as reflecting an average offense facing an average defense.
These estimates apply not to an average offense but to a randomly drawn offense from the dataset of plays.
Because good offenses run more plays and score more points, these estimates are size biased; they reflect an offense whose quality is a weighted average of offenses tilted towards good offenses.
The same logic applies to defense.
Selection biases of this sort appear in other estimation tasks across sports analytics.
In soccer, great players take more shots and score more goals.
Thus, standard expected goals models don't apply to average players \citep{davis2024biasesexpectedgoalsmodels}.
In basketball, great shooters take more shots and score more points, particularly three-pointers.
Thus, naive shot probability models overestimate the probability an average player makes a shot from certain locations.
Across sports analytics, failing to adjust for team quality produces bias and erodes the quality of models.
This issue is pervasive and hasn't been thoroughly addressed in sports analytics literature.

Adjusting for team quality further complicates the task of estimating $\ep$.
There are so many dimensions of team quality -- such as offensive quality, defensive quality, quarterback quality, or even kicker quality -- and many ways to quantify these dimensions.
Compounded with an already large space of game-states, measures of team quality balloon the dimensionality of the problem.
The number of variables explodes and there are not that many observations.
On one hand, a statistician may deal with this by fitting a statistical model, whose parameters facilitate borrowing strength across similar game-states. 
But, those models underfit, as it is difficult to fit a sufficiently rich statistical model that captures important interactions between so many variables.
On the other hand, a modern practitioner may deal with this using machine learning, whose flexibility can capture important non-linearities and interactions.
But, machine learning overfits, producing artifacts that contradict fundamental principles.

The strong dependence structure of the data, which significantly cuts down the number of independent observations, makes it even easier for machine learning tools to overfit.
Analysts haven't accounted for, let alone acknowledged, the dependency structure; they implicitly assume that the observational play-by-play football dataset consists entirely of independent plays.
This is not true because all plays in an epoch share the same outcome.
Ignoring the dependency structure potentially produces biased estimates and complicates the task of trying to do uncertainty quantification.   

Uncertainty estimates are crucial to player evaluation, but standard $\ep$ approaches do not provide confidence intervals to accompany their point estimates.
Thus, it is not possible to discern whether differences in $\epa$ between players are due to skill or random chance.

In this work, we deal with each of these problems in turn.
First, to mitigate selection bias, we adjust for team quality.
Second, to reduce overfitting, we devise a novel method to regularize the fits of flexible, nonparametric models using synthetic data generated from simpler, parametric models. 
Third, we carefully examine the extent to which ignoring the dependency structure affects our ability to quantify uncertainty and we present several alternatives.
In particular, we find that by bootstrapping an appropriately weighted $\xgb$ model, we can adequately quantify uncertainty.

Analysts who use machine learning tools in lieu of traditional statistical models often encounter the same problems we face here. 
Quantifying uncertainty using machine learning when facing size bias or selection bias, overfitting, and strongly correlated data presents a serious challenge, particularly when there is much less data than you think.
In this work, we provide a case study exemplifying ways to think about and deal with these challenges. 
While we develop an approach for our particular football application, resulting in a more refined expected points model, the ideas and methods in this work generalize to other applications.

The remainder of this paper is organized as follows.
In Section~\ref{sec:problems_trad_ep} we provide a careful review of the statistical assumptions and modeling choices made by standard $\ep$ models.
In Section~\ref{sec:accounting_dep_struct} we devise expected points models that account for the dependency structure of historical football data and fully capture uncertainty in drive outcomes.
In Section~\ref{sec:smoothing_xgb} we introduce the use of a catalytic prior to smooth overfitting in blackbox machine learning models.
Then, in Section~\ref{sec:player_eval_results} we knit together the lessons learned in this paper to evaluate NFL offenses and quarterbacks. 
In Section~\ref{sec:discussion} we conclude.

\section{Examining assumptions and modeling choices made by traditional by $\ep$ models}
\label{sec:problems_trad_ep}

In this section, we provide a careful review of the statistical assumptions and modeling choices made by standard $\ep$ models.

\subsection{Epoch outcome versus drive outcome}
\label{sec:epoch_vs_drive_ep}

In this section, we discuss why throughout this paper we use $\ep$ fit from a dataset grouped into drives rather than epochs.
We begin by introducing some football terminology. 
On one hand, a \textit{drive} is a series of contiguous plays featuring one offensive team, ending with a score or a change in possession.  
The outcome $y$ of a drive has five potential outcomes, each with an associated net points value shown in parenthesis:  touchdown (7), field goal (3), no score (0, e.g.~turnover, touchback punt, etc.), opposing team safety (-2), and opposing team touchdown (-7, e.g.~turnover returned for a touchdown).  
On the other hand, an \textit{epoch} is a series of contiguous plays resulting in the same next scoring event in the half.
The outcome $y$ of an epoch has seven potential outcomes, each with an associated net points value shown in parenthesis: touchdown (7), field goal (3), safety (2), no score (0), opponent safety (-2), opponent field goal (-3), and opponent touchdown (-7).
Today, both epoch and drive $\ep$ are typically defined by a weighted sum of the probabilities of their respective outcomes,
\begin{equation}
\label{eqn:p_to_ep}
\ep(\bx) = \sum_{k} \pts(k) \cdot \P(y=k|\bx),
\end{equation}
where $\bx$ denotes the game-state, $\pts(k)$ denotes the net points associated with outcome $k$, and $\P(y=k|\bx)$ is estimated from data.
Each epoch consists of at least one drive, and on average consists of $1.88$ drives with a long right tail (e.g., $10\%$ of epochs consist of at least $4$ drives).  

Football analysts traditionally define $\ep$ as expected points of the epoch rather than expected points of the drive because the ``no score’’ outcome of a drive could be favorable or unfavorable depending on the game-state $\bx$.
For instance, a turnover 96 yards from the opponent’s endzone is much worse than a turnover 4 yards from the opponent’s endzone.
Drive $\ep$ doesn't capture the difference between these two outcomes because each drive results in ``no score,'' whereas epoch $\ep$ captures this difference since it is based on the next scoring event.

Although drive $\ep$ and epoch $\ep$ may differ, drive $\epa$ and epoch $\epa$ similarly value the success of a play (recall $\epa$, or expected points added, is the difference in $\ep$ between the end and the beginning of a play).
For example, a turnover 96 yards from the opponent's endzone is viewed similarly by drive $\epa$ and epoch $\epa$ because after the turnover, the opposing team has extremely good field position.
Because applications  of $\ep$ are based on changes in $\ep$ between plays (i.e., $\epa$), not on the raw $\ep$ of a single play, we don't view the aforementioned advantage of epoch $\ep$ as a strong reason to favor epoch $\ep$ over drive $\ep$ in evaluating players via $\epa$.

Furthermore, drive $\ep$ has substantial advantages over epoch $\ep$.  
Each epoch consists of at least one drive, usually consisting of multiple drives featuring multiple changes in possession before a scoring event.  
Thus, a single play on average has a much larger impact on the outcome of a drive than on the outcome of an epoch.  
To see this, imagine an epoch in which possession goes back and forth, say, six times before a score.  
The plays in the first four (and potentially five) drives have a negligible impact on the outcome of that epoch.  
Ultimately, it is the successful play of the sixth and final drive that leads to the epoch outcome. 
All of the plays in each of those six drives, however, have a tangible impact on the outcome of each respective drive. 
The outcome of a drive is much more immediate than the outcome of an epoch, which is significantly noisier, and drive $\ep$ has a much larger effective sample size than epoch $\ep$.  

For these reasons, in this paper we use drive $\ep$ rather than epoch $\ep$.\footnote{
    The qualitative results of this paper still hold for epoch $\ep$.
}

\subsection{Selection bias}
\label{sec:selection_bias}

Because American football analysts want to evaluate players using $\epa$, they are interested in a context-neutral expected points function that assumes an offense faces an average defense and a defense faces an average offense.
For instance, \citet{Romer06dofirms} explains that his $\ep$ model represents the ``expected long-run value ... of the difference between the points scored by the team with the ball and its opponent when the two teams are evenly matched, average NFL teams.''
To create context-neutral $\ep$ models, analysts have not included measures of team quality in the game-state vector $\bx$.

\begin{figure}[hbt!]
    \centering{}
    \subfloat[\centering ]{
        {\includegraphics[width=0.3\textwidth]{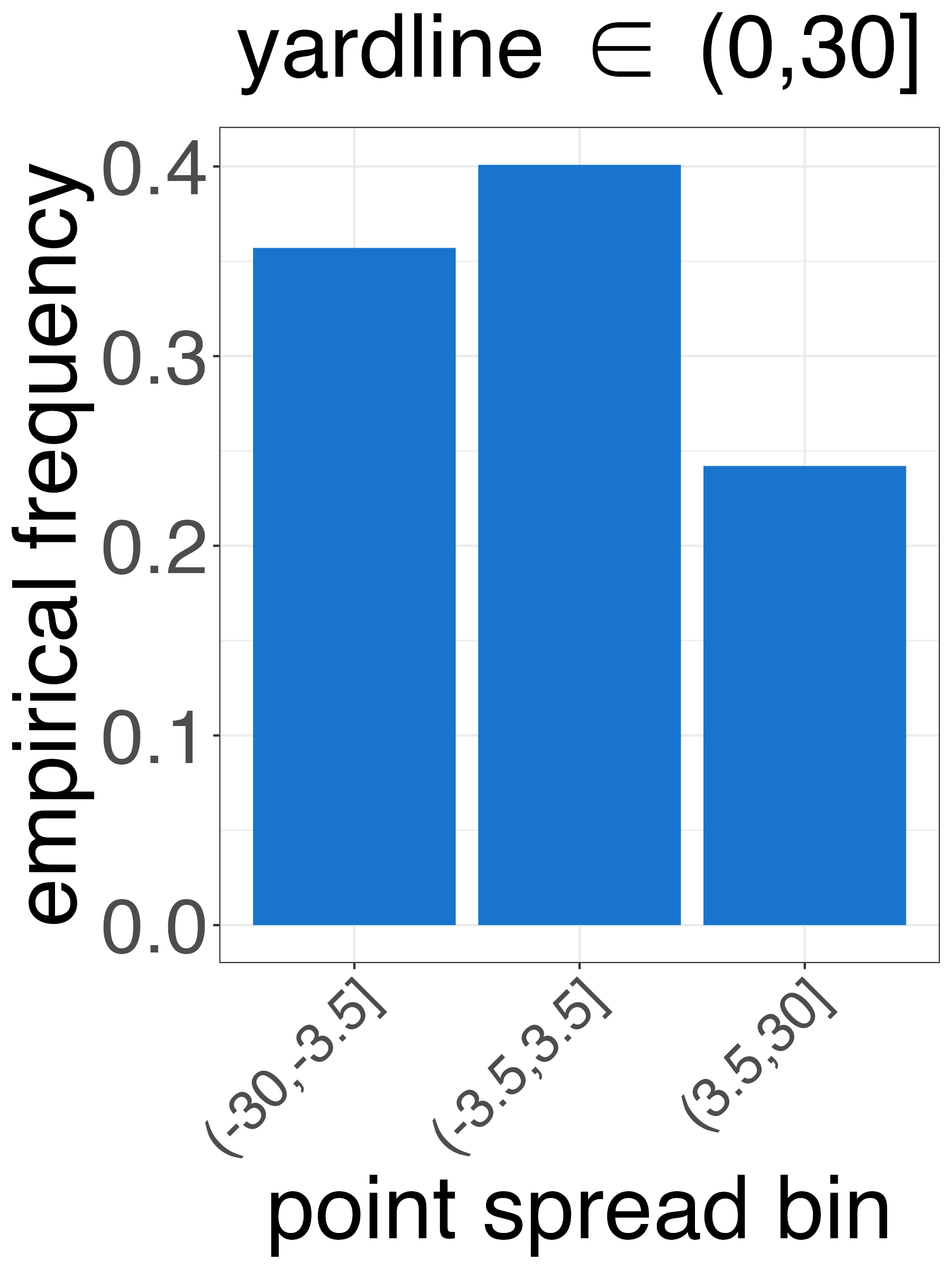}}
        \label{fig:plot_selection_bias1A}
    }
    \qquad
    \subfloat[\centering ]{
        {\includegraphics[width=0.3\textwidth]{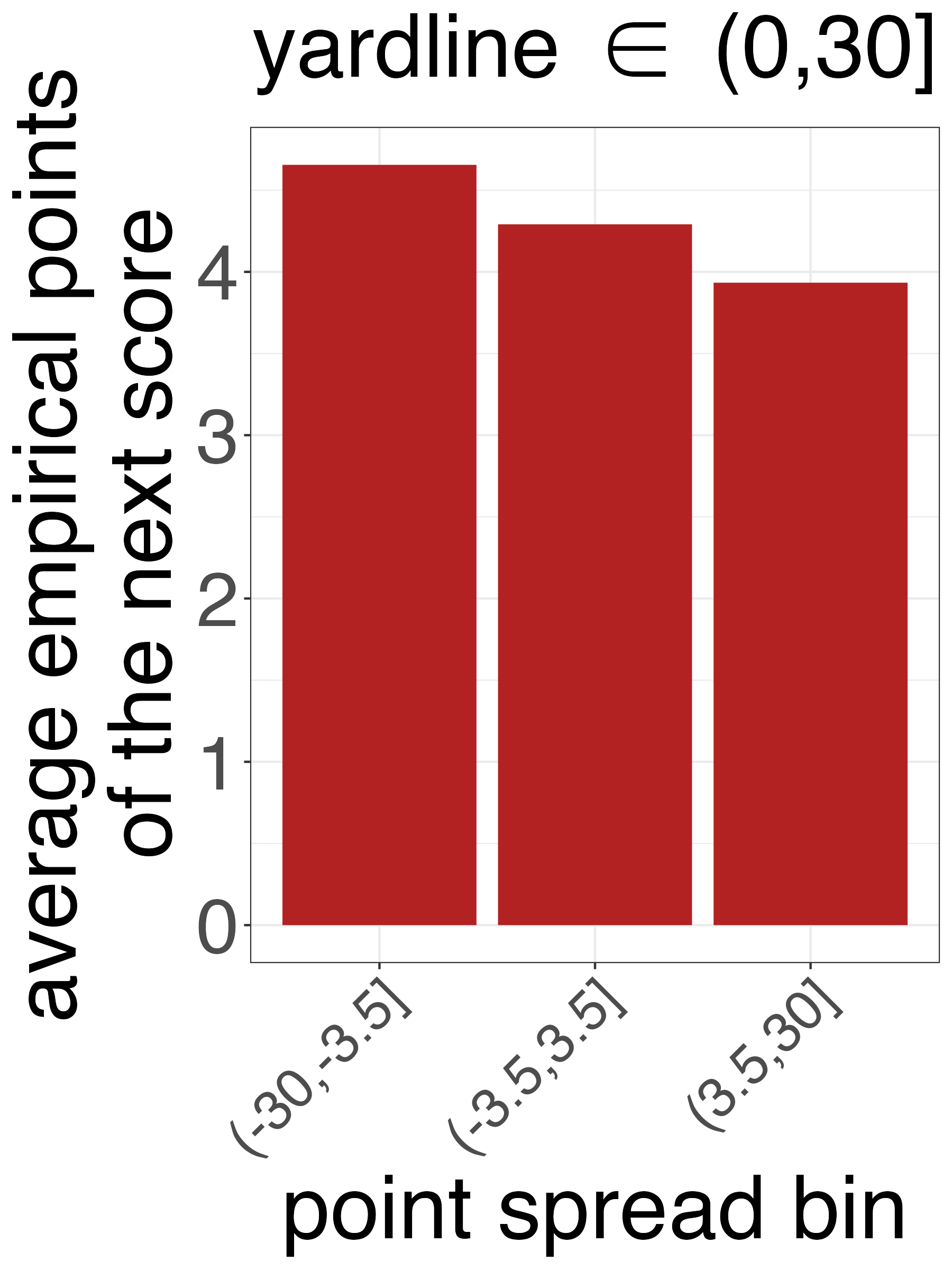}}
        \label{fig:plot_selection_bias1B}
    }
    \caption{
        Good teams have more plays (a) and score more points (b) conditional on being near the opposing team's endzone.
        Figure~(a): a higher proportion of plays feature good teams (negative point spreads) than bad teams (positive point spreads).
        Figure~(b): the average empirical net points of the drive is higher for good teams. 
    }
    \label{fig:plot_selection_bias}
\end{figure}

Failing to adjust for team quality causes selection bias.
Suppose that good teams have a pre-game point spread strictly below $-3$ and bad teams strictly above $+3$.
Averaged over all plays, good teams run $32\%$ of plays and bad teams run $26\%$.
Also, good teams average $0.7$ more points per drive than bad teams.
We visualize these imbalances near the opponent's endzone in Figure~\ref{fig:plot_selection_bias}.
Because good teams run more plays and score more points, standard play-level regression $\ep$ models are sized biased.
Those models fit expected points not for an average team but for a randomly drawn team from the dataset of plays, whose quality is a weighted average of teams tilted towards good teams.
Thus, they overestimate expected points for average teams.

To mitigate this selection bias, we can include measures of team quality (e.g., point spread) as covariates.
Then, to get $\ep$ for average teams, we evaluate an $\ep$ model using average team qualities (e.g., 0 point spread).
We employ this approach in Section~\ref{sec:accounting_dep_struct}.

\subsection{The dependency structure of observational football data}
\label{sec:dep_structure}

The traditional formulation of $\ep$ overlooks the dependency structure of the data, which could correspond to a mis-specified model and complicates the task of trying to do uncertainty quantification.  
Formally, in our dataset of observed football plays, let $i$ index the drive and $j$ index the play within each drive.  
As before, let $y$ denote the drive outcome and $\bx$ denote the game-state.
Then, a multinomial logistic model for the probability of each drive outcome $k$ is specified by
\begin{equation}
\label{eqn:mlogistic_model_ij}
\log\bigg(\frac{\P(y_{ij} = k|\bx_{ij})}{\P(y_{ij} = \text{No Score}|\bx_{ij})}\bigg) = f_k(\bx_{ij}).
\end{equation}
As per Section~\ref{sec:epoch_vs_drive_ep}, we use drive $\ep$ throughout this paper even though traditional $\ep$ models actually use epoch $\ep$.
This model is mis-specified because all plays $j$ in the same drive $i$ share the same outcome, $y_{ij} \equiv y_i$.  
There is one outcome $y_i$ generated per drive, but model~\eqref{eqn:mlogistic_model_ij} assumes a separate multinomial outcome $y_{ij}$ is generated per play.  
This potentially produces a biased model because some drives consist of many more plays, each of which contributes less to the ultimate drive outcome than plays from smaller drives.
It also exacerbates the selection bias issue from Section~\ref{sec:selection_bias}.
Play-level regression models that don't adjust for team quality fit expected points for a randomly drawn team, which is above average since good teams run more plays.
Conversely, models that weigh each drive equally, rather than each play equally, do not overweigh good teams since drives alternate between teams.

\subsection{Capturing uncertainty in the outcome of a drive}
\label{sec:capture_uncertainty_in_drive_outcome}

A key question arises: can we get satisfactory uncertainty quantification from a traditional $\ep$ model~\eqref{eqn:mlogistic_model_ij} that overlooks the dependency structure of the data?
Suppose an analyst produces a vector of estimated drive outcome probabilities for each game-state $\bx$,
\begin{equation}
  \bphat(\bx) = (\phat_{\text{touchdown}}(\bx), \ \phat_{\text{field goal}}(\bx), \ \phat_{\text{no score}}(\bx), \ \phat_{\text{opp. safety}}(\bx), \ \phat_{\text{opp. touchdown}}(\bx)).
  \label{eqn:p_vec}
\end{equation}
We'd like to know whether the probability estimates are ``good'' before using them to value players or recommend strategic decisions.
Since the ``true’’ probabilities are unobservable, we'd like to know whether our estimated probabilities~\eqref{eqn:p_vec} adequately capture uncertainty in the outcome of a drive.

One way to convey uncertainty in the outcome of a drive using $\bx \mapsto \bphat(\bx)$ is a prediction set: given the game-state, what set of outcomes is likely to occur?
Formally, what is the smallest set of outcomes that collectively contains at least 95\% of the probability mass?  
A model that successfully captures uncertainty in the outcome of a drive should produce prediction sets that achieve adequate marginal coverage.  
In other words, a 95\% prediction set should, over average, contain (or, cover) the true outcome for 95\% of plays.

Marginal calibration is a necessary condition for a probability model to be good, and \citet{nflFastREPxgb} found that his traditional $\ep$ model is calibrated. 
But calibration is not a sufficient condition for a model to be ``good’’: a model can be calibrated but wildly inaccurate.\footnote{
    The weather is a prime example.  Consider Philadelphia, where it rains about 127 days per year on average.  A constant daily rain probability estimate of 127/365 = 0.348 is calibrated but is a terrible predictor, as rainy days are likely to appear in clumps and sunny days are likely to appear in clumps.
}  
Model accuracy is also not enough to tell us whether probability estimates are ``good.’’  
Since the outcome of a drive is so noisy, model accuracy metrics like $\logloss$ or $\rmse$ will be large no matter what model we use.
Model accuracy is useful for model comparison (e.g., we’d rather use a more accurate model, ceteris paribus).

We find that traditional $\ep$ models á la Formula~\eqref{eqn:mlogistic_model_ij} produce prediction sets that do not achieve adequate marginal coverage (see Section~\ref{sec:accounting_dep_struct}).   
As those models produce undercovered prediction sets, they do not adequately capture uncertainty in the outcome of a drive.
Hence, in Section~\ref{sec:accounting_dep_struct} we devise $\ep$ models that account for the dependency structure of observational football data and satisfactorily capture uncertainty in the outcome of a drive.

\subsection{Artifacts of overfitting in $\xgb$ models}
\label{sec:overfitting_xgb}

Football analysts have embraced recent advances in machine learning and were quick to incorporate $\xgb$ into the $\ep$ framework \citep{nflFastREPxgb}.
In theory, models like $\xgb$ are ideal for estimating $\ep$ since the function $\bx \mapsto \bphat(\bx)$ likely includes nonlinearities and complex interactions between variables.
In practice, the flexible nature of blackbox machine learning models makes them prone to overfitting.
Although our play-by-play dataset consisting of about $500,000$ plays (see Section~\ref{sec:data}) is ostensibly massive, the space of game-states is extremely large (recall Section~\ref{sec:intro}), making it difficult to capture complex trends amongst the covariates.
Adjusting for team quality to mitigate selection bias further balloons the size of the state space.
The dependency structure further reduces the effective sample size, exacerbating the problem.

\begin{figure}[hbt!]
    \centering{}
    \subfloat[\centering ]{
        {\includegraphics[width=0.5\textwidth]{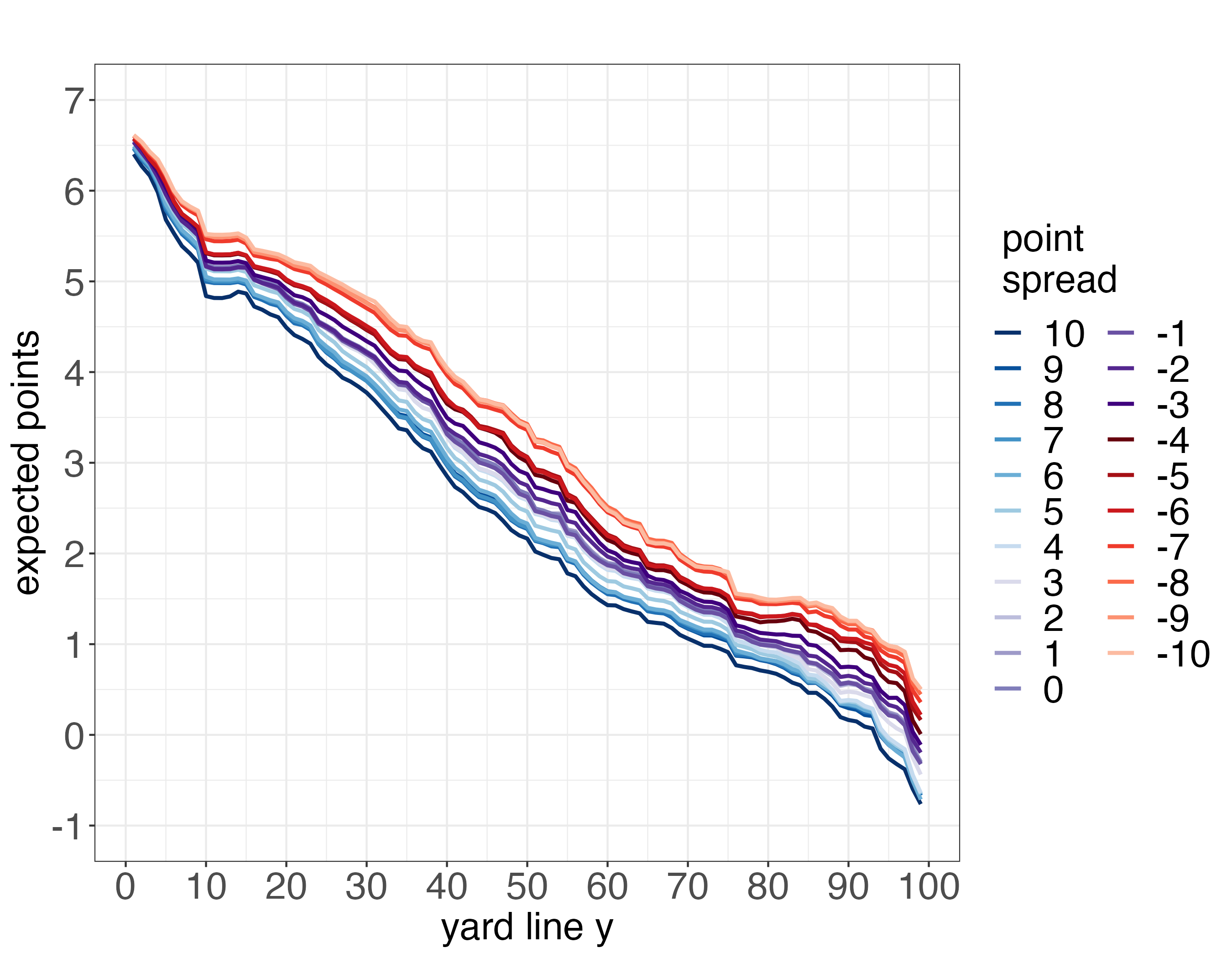}}
        \label{fig:wxgb_tq_L}
    }
    \subfloat[\centering ]{
        {\includegraphics[width=0.5\textwidth]{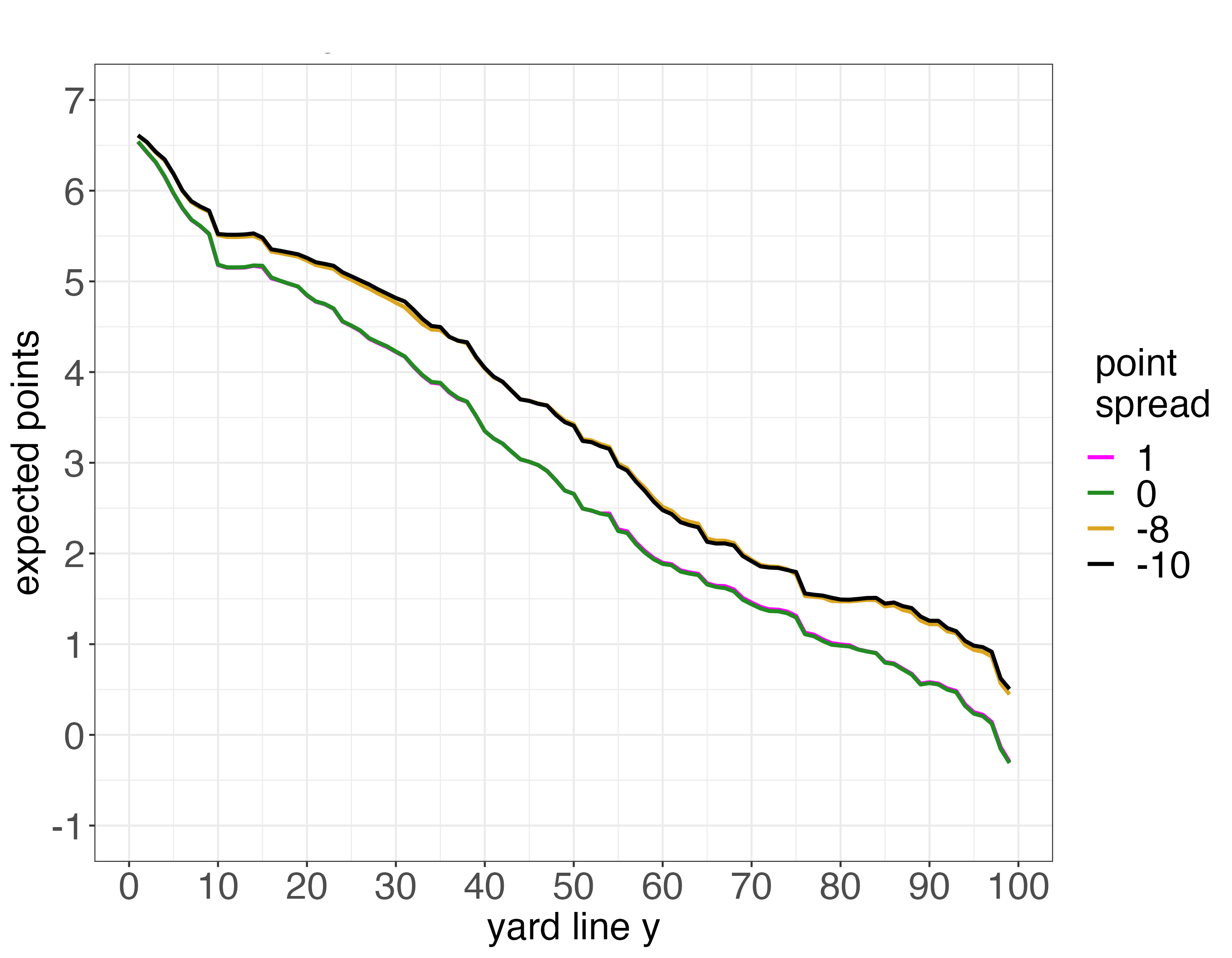}}
        \label{fig:wxgb_tq_R}
    }
    \caption{
        $\ephat$ ($y$-axis) according to weighted multiomial $\xgb$ as a function of yardline ($x$-axis) and pre-game point spread relative to the team with possession (color), holding the other covariates fixed.  In (b) we see artifacts of overfitting, that $\ephat$ is not monotonic decreasing in point spread even though it should be.
    }
    \label{fig:wxgb_tq}
\end{figure}

We visualize an example of overfitting in Figure~\ref{fig:wxgb_tq}, which views $\ephat$ as a function of yardline and point spread.
Figure~\ref{fig:wxgb_tq_L} shows that the model finds that decreasing the point spread (or, increasing the difference in team quality between the offensive and defensive team) is associated with increasing expected points.
This matches the basic intuition that we should expect better teams to score more points.
But we also see some counter-intuitive results with respect to point spread.
In Figure~\ref{fig:wxgb_tq_R} we see that $\ephat$ for a $-8$ point spread is (slightly) larger than $\ephat$ for a $-10$ point spread for some yardlines (the orange line is above the black line).
We also see that $\ephat$ for a $+1$ point spread is (slightly) larger than $\ephat$ for an even point spread for some yardlines (the pink line is above the green line).
These examples defy our intuition and are plausibly artifacts.
In light of these findings, we explore smoothing such artifacts of overfitting in Section~\ref{sec:smoothing_xgb}.

\section{Accounting for the dependency structure of historical football data}\label{sec:accounting_dep_struct}

In this section, we devise expected points models that account for the dependency structure of historical football data and more thoroughly capture uncertainty in drive outcomes.

\subsection{Data}\label{sec:data}

We access every NFL play from 2010 to 2022 using the $\R$ package $\nflfastr$ \citep{nflFastR}, yielding a dataset of $491,993$ plays, with $73,514$ drives and $39,083$ epochs.
Each play includes variables that describe the context of the play, which are relevant to estimating expected points, such as the yardline, down (categorical), yards to go, half seconds remaining, era, offensive team timeouts remaining, defensive team timeouts remaining, score differential, pre-game point spread relative to the team with possession, and drive outcome.
The code for this study, which includes code to scrape the dataset, is publicly available on Github.\footnote{
    \url{https://github.com/snoopryan123/expected_points_nfl}
} 

\subsection{Models that account for the dependency structure}\label{sec:models_dep_structure}

Traditional expected points models haven't considered the dependency structure of observational football data.
$\ep$ is traditionally framed as a regression problem estimated at the level of plays, treating the outcome of each play as an independent draw (recall model~\eqref{eqn:mlogistic_model_ij}).
Formally, traditional drive outcome probability models minimize the $\logloss$ of the training set, weighing each play equally, 
\begin{equation}
\argmin_{p} \left\{ -\sum_{i=1}^{\Ndrives} \sum_{j=1}^{N_i} \sum_{k=1}^{K} \ind{y_{i}=k} \cdot \log p(k|\bx_{ij}) \right\}.
\label{eqn:loss_EP_trad}
\end{equation}
Here, $i$ indexes the drive (from 1 to $\Ndrives$), $j$ indexes the play within the drive (from 1 to $N_i$, the number of plays in drive $i$), $k$ indexes all possible outcomes (from 1 to $K=5$), $\bx_{ij}$ denotes the game-state, and $y_{i}$ denotes the drive outcome. 
\citet{nflWar} use multinomial logistic regression and \citet{nflFastREPxgb} uses $\xgb$ to solve this optimization problem. 

But, all plays within the same drive share the same outcome of the drive.
To account for this dependency structure, we devise the \textit{averaged subsample} model.
$M=100$ times we uniformly draw one play per drive, which yields a subsample consisting of plays having independent outcomes. From each subsample we fit a model (e.g., $\xgb)$ and then average the $M$ models.  
The averaged subsample model is trained by minimizing the averaged $\logloss$ across the $M$ randomly subsampled training sets,
\begin{equation}
\argmin_{p} \left\{ -\frac{1}{M}\sum_{m=1}^{M}\sum_{i=1}^{N_{\text{drives}}} \sum_{j=1}^{N_i} I_{mij} \sum_{k=1}^{K} \ind{y_{i}=k} \cdot \log p(k|\bx_{ij}) \right\},
\label{eqn:loss_EP_avgsub}
\end{equation}
where
\begin{equation}
I_{mij} = \ind{\text{play $j$ is sampled from drive $i$ in train set subsample $m$}}.
\end{equation}
Although the averaged subsample model accounts for the dependency structure, it is extremely computationally intensive, particularly when we bootstrap it later in Section~\ref{sec:boot_models}.

The averaged subsample model is a random model since it is trained on $M$ randomly drawn subsamples of the training set.
The expected value of the averaged subsample model is deterministic and accounts for the dependency structure.
But it is computationally infeasible to fit the expected value of the averaged subsample model. 
We instead approximate this model by switching the order of expectation and minimization,
\begin{align}
& \bE\bigg[ \argmin_{p} \left\{ -\frac{1}{M}\sum_{m=1}^{M}\sum_{i=1}^{N_{\text{drives}}} \sum_{j=1}^{N_i} I_{mij}\sum_{k=1}^{K} \ind{y_{i}=k} \cdot \log p(k|\bx_{ij}) \right\} \bigg]  \\
\approx & \argmin_{p} \left\{ -\frac{1}{M}\sum_{m=1}^{M}\sum_{i=1}^{N_{\text{drives}}} \sum_{j=1}^{N_i} \bE[I_{mij}] \sum_{k=1}^{K} \ind{y_{i}=k} \cdot \log p(k|\bx_{ij}) \right\} \\
= & \argmin_{p} \left\{ - \sum_{i=1}^{N_{\text{drives}}} \sum_{j=1}^{N_i} w_{ij} \sum_{k=1}^{K} \ind{y_{i}=k} \cdot \log p(k|\bx_{ij}) \right\},
\label{eqn:loss_EP_avgsub}
\end{align}
where 
\begin{equation}
w_{ij} = \P(\text{play $j$ is sampled from drive $i$ in train set subsample $m$}) = \frac{1}{N_i}.
\end{equation}
In other words, we fit a model that minimizes weighted $\logloss$, weighing each play by the inverse number of plays in the drive it appears (e.g., weighted $\xgb$).
This upweighs plays from smaller drives and downweighs plays from larger drives, which intuitively makes sense because plays from smaller drives more directly impact the outcome of the drive.
This model has the same computational cost as traditional $\ep$ models (e.g., unweighted $\xgb$).

\subsection{Model evaluation}\label{sec:model_eval}

Each outcome probability model $\bphat$ yields an associated expected points model $\ephat$ (via Formula~\eqref{eqn:p_to_ep}) and a prediction set $\predsetp$ (the smallest set of outcomes that collectively contains at least 95\% of the estimated probability mass).
We want to see whether accounting for the dependency structure improves model accuracy, and we want to learn whether these models adequately account for uncertainty in the outcome of a drive.

Traditional evaluations of expected points models haven’t considered the dependency structure: compute accuracy, calibration, or coverage on an out-of-sample test set, weighing all plays equally.
This, however, doesn't match the use case of $\ep$ models:  we're watching a football game, a play at a certain game-state arises, and we want to predict the outcome of the drive as best as possible.  
Mimicking this use case, we want a model to perform as best as possible on a play that is {randomly drawn from a held-out drive}.
So, we evaluate $\ep$ models on a subsample of the out-of-sample test set $\Dtest$ formed by uniformly drawing one play per drive.
To de-noise the evaluation, we average the evaluation metrics across $\Mtest=100$ test set subsamples.

Formally, we evaluate the accuracy of $\ephat$ on the $m^{th}$ subsampled test set by the $\rmse$,
\begin{equation}
\rmse_m(\ephat, \Dtest) = \sqrt{ \frac{1}{\Ndrives} \sum_{i=1}^{\Ndrives} \sum_{j=1}^{N_i} I_{mij} \cdot (\ephat(\bx_{ij}) - \pts(y_{i}))^2},
\label{eqn:test_rmse_m}
\end{equation}
where $\pts(y_{i})$ is the net points value associated with the drive outcome $y_{i}$ and
\begin{align}
I_{mij} = \ind{\text{play $j$ is sampled from drive $i$ in test set subsample $m$}}.
\label{eqn:I_mij_test}
\end{align}
We then average $\rmse$ across the $\Mtest$ subsampled test sets,
\begin{equation}
\rmse(\ephat, \Dtest) = \frac{1}{\Mtest} \sum_{m=1}^{\Mtest} \rmse_m(\ephat, \Dtest).
\label{eqn:test_rmse}
\end{equation}
We also compute its standard error,
\begin{equation}
\SE(\rmse(\ephat, \Dtest)) = \SD(\{ \rmse_m(\ephat, \Dtest) : 1 \leq m \leq \Mtest \}) / \sqrt{\Mtest},
\label{eqn:test_rmse_SE}
\end{equation}
where $\SD$ denotes the sample standard deviation.

Similarly, we evaluate the accuracy of $\bphat$ by the average $\logloss$ across the $\Mtest$ subsampled test sets,
\begin{equation}
\logloss(\bphat, \Dtest) = - \frac{1}{\Mtest} \sum_{m=1}^{\Mtest} \frac{1}{\Ndrives} \sum_{i=1}^{\Ndrives} \sum_{j=1}^{N_i} I_{mij} \sum_{k=1}^{K} \ind{y_{i}=k} \cdot \log \phat(k|\bx_{ij}),
\label{eqn:test_logloss}
\end{equation}
and its standard error.
Finally, we evaluate the coverage of $\predsetp$ by the average marginal coverage across the $\Mtest$ subsampled test sets,
\begin{equation}
\covg(\bphat, \Dtest, \alpha) = \frac{1}{\Mtest} \sum_{m=1}^{\Mtest} \frac{1}{\Ndrives} \sum_{i=1}^{\Ndrives} \sum_{j=1}^{N_i} I_{mij} \cdot \ind{y_{i} \in \predsetp(\bx_{ij}, \alpha)},
\label{eqn:test_covg}
\end{equation}
and its standard error.
Here, $\alpha = 95\%$ refers to a $95\%$ prediction set.


Now, we evaluate variants of a base $\xgb$ model that either do or do not account for the dependency structure -- unweighted $\xgb$, averaged subsampled $\xgb$, and weighted $\xgb$ -- using the aforementioned metrics.
Each of these models drive outcome probability as a function of yardline, down (categorical), yards to go, half seconds remaining, era, offensive team timeouts remaining, defensive team timeouts remaining, score differential, and pre-game point spread relative to the team with possession.
We include point spread to adjust for team quality, which mitigates the selection bias discussed in Section~\ref{sec:selection_bias}.

The full out-of-sample test set $\Dtest$ consists of all plays from a randomly sampled $25\%$ of the drives from the full dataset. 
To tune the $\xgb$ model, we split the training set in half by randomly sampling $50\%$ of the drives in order to preserve the drive dependency structure.
This forms a hyperparameter validation and training set.
We then tune our $\xgb$ models by searching over a grid of parameter combinations, selecting the parameters that minimize validation $\logloss$ \citep{BaldwinWP}.
We tune the weighted $\xgb$ separately from the unweighted $\xgb$, and we use the unweighted $\xgb$'s tuned parameters for the averaged subsampled $\xgb$.

In Table~\ref{table:model_eval} we display the results of our model evaluation.
According to both $\rmse$ and $\logloss$, unweighted $\xgb$ is less accurate than both weighted $\xgb$ and averaged subsampled $\xgb$.
Those two models have essentially the same accuracy.
Although the improvement in accuracy is slight, accounting for the dependency structure improves model accuracy nonetheless.
We suspect that accounting for the dependency structure improves accuracy because plays in shorter drives more directly impact the outcome of the drive than plays in longer drives.

On the other hand, prediction sets from unweighted $\xgb$ achieve slightly higher coverage than those from the other $\xgb$ variants.
This curiosity is overshadowed by the observation that prediction sets formed from each of these models are undercovered: they achieve about $85\%$ coverage even though they are supposed to achieve $95\%$ coverage.
In other words, nominally $95\%$ prediction sets formed from the point estimates of a lone drive outcome probability model don't account for the full uncertainty in the outcome of a drive.
This suggests that we need to amend our models to fully account for this uncertainty.

\begin{table}[htb!]
\centering
\begin{tabular}{llll} \hline
  model $\bphat$ & $\rmse(\ephat, \Dtest)$ & $\logloss(\bphat, \Dtest)$ & $\covg(\bphat, \Dtest, 95\%)$ \\ \hline 
  weighted $\xgb$ & $2.593 \pm 0.0017$ & $0.7506 \pm 0.0006$ & $0.834 \pm 0.0004$ \\ 
  averaged subsampled $\xgb$ & $2.593 \pm 0.0017$ & $0.7521 \pm 0.0006$ & $0.841 \pm 0.0004$ \\ 
  unweighted $\xgb$ & $2.618 \pm 0.0014$ & $0.7670 \pm 0.0005$ & $0.861 \pm 0.0004$ \\ \hline
\end{tabular}
\caption{
Model evaluation results for variants of an $\xgb$ drive outcome probability model (plus-minus twice the standard error).
}
\label{table:model_eval}
\end{table}


\subsection{Bootstrapped models}\label{sec:boot_models}

The aforementioned models don't fully capture uncertainty in the outcome of a drive.
These approaches to estimating $\ep$ produce undercovered prediction sets because they don’t account for sampling uncertainty, or uncertainty that arises from estimating a quantity from a finite noisy dataset.  
To understand,
suppose there is some ``true’’ probability model generating the data.  
If we could re-run the recent history of football, by the randomness inherent in generating the outcome $y$ from $\bx$ the resulting play-by-play dataset would be different.  
Thus, the model $\bphat$ estimated from the dataset would be different.  
It is important to quantify this source of uncertainty in $\ep$ estimates because it impacts how we value players and how we recommend strategic decisions.  
For instance, we'd like to know how granularly to trust differences in estimated $\epa$ across players.  
Is the difference between, say, Mahomes and Allen’s $\epa$ per play in 2022 due to random chance or systematic skill? 
And is an estimated edge in $\ep$ by making a certain strategic decision ``real'' or due to noise?
Bootstrapping is a natural choice to capture this source of uncertainty since it is a nonparametric method that doesn’t assume a functional parametric form for the probability model.
This works well for blackbox machine learning models like $\xgb$. 

On a high level, we bootstrap $B$ datasets $\{(\bX^{(b)}, \by^{(b)}) : 1 \leq b \leq B\}$ from the original training dataset $(\bX,\by)$ and fit a drive outcome probability model $\bphat^{(b)}$ from each dataset.
The standard (i.i.d.) bootstrap generates a bootstrapped dataset by resampling plays (rows) uniformly with replacement.  
The cluster bootstrap generates a bootstrapped dataset by resampling drives (clusters) uniformly with replacement.  
The latter accounts for the dependency structure and the former does not.

To assess the efficacy of the bootstrap in capturing uncertainty in the outcome of a drive, we consider coverage of the drive outcome by bootstrap-informed prediction sets.
Creating a prediction set from $B$ bootstrapped drive outcome probability models $\{\bphat^{(b)}\}_{b=1}^{B}$ is not as straightforward as creating a prediction set from just one model $\bphat$.
Before, we simply used the smallest set of outcomes whose estimated summed probability according to $\bphat$ exceeds 95\%.
To extend this to $B$ probability models, we generate an outcome $\hat{y}_{mj}^{(b)}$ for each play $j$ in each subsampled test set $m$ from each bootstrapped drive outcome probability model $\bphat^{(b)}$.
The prediction set for the $j^{th}$ play in the $m^{th}$ subsampled test set is then formed from the empirical distribution of $\{\hat{y}_{mj}^{(b)}\}_{b=1}^{B}$.
Specifically, it is the smallest set of outcomes whose estimated summed probability according to that empirical distribution exceeds 95\%.
Then, the coverage of our bootstrapped prediction intervals is, like before, the average marginal coverage across the $\Mtest$ subsampled test sets,
\begin{equation}
\bootcovg(\{\bphat^{(b)}\}_{b=1}^{B}, \Dtest, \alpha) = \frac{1}{\Mtest} \sum_{m=1}^{\Mtest} \frac{1}{\Ndrives} \sum_{i=1}^{\Ndrives} \sum_{j=1}^{N_i} I_{mij} \cdot \ind{y_{i} \in \predsetpboot(\bx_{ij}, \alpha) }.
\label{eqn:test_bootcovg}
\end{equation}
$I_{mij}$, as in Formula~\eqref{eqn:I_mij_test}, indicates whether the $j^{th}$ play of the $i^{th}$ drive is in the $m^{th}$ subsampled test set.
Here, $\alpha = 95\%$ refers to a $95\%$ prediction set.
We also compute the standard error as before.

In Table~\ref{table:model_eval_boot} we display the results of our bootstrap evaluation with $B=100$.
None of the bootstrap variants yield prediction sets that achieve significantly higher coverage than the others.
Further, each of the bootstrap variants yield prediction sets that achieve about $95\%$ marginal coverage as desired.
This indicates that we need bootstrapping, or something else beyond just the point estimates of a lone drive outcome probability model, to capture the full amount of uncertainty in the outcome of a drive.

\begin{table}[htb!]
\centering
\begin{tabular}{lll} \hline
  model $\bphat$ & bootstrap & value ($\pm 2\cdot\SE$) \\ \hline
  weighted $\xgb$ & cluster bootstrap & $0.956 \pm 0.016$ \\ 
  averaged subsampled $\xgb$ & cluster bootstrap & $0.957 \pm 0.016$  \\
  unweighted $\xgb$ & standard i.i.d. bootstrap & $0.963 \pm 0.013$ \\ \hline
\end{tabular}
\caption{Bootstrap evaluation results according to $\bootcovg(\{\bphat^{(b)}\}_{b=1}^{B}, \Dtest, \alpha=95\%)$. }
\label{table:model_eval_boot}
\end{table}

In Section~\ref{sec:models_dep_structure} we devised the averaged subsampled $\xgb$ and weighted $\xgb$ models which, unlike traditional $\ep$ models such as unweighted $\xgb$, account for the dependency structure of observational football data.
In Section~\ref{sec:model_eval} we found that the models that account for the dependency structure are more accurate than models that do not, but none of those models produce prediction sets that achieve adequate marginal coverage.
Thus, although accounting for the dependency structure improves the model, it still doesn't fully capture uncertainty in the outcome of a drive.
To account for an additional source of uncertainty, sampling uncertainty, in this Section~\ref{sec:boot_models} we constructed bootstrapped $\ep$ models.
Each of the bootstrap-informed prediction sets achieves adequate marginal coverage.
This suggests that sampling uncertainty, not the dependency structure, was the primary culprit responsible for the undercoverage of traditional $\ep$ models.
Even though traditional models~\eqref{eqn:mlogistic_model_ij} are mis-specified, they are adequate if they are coupled with bootstrapping.

\section{Smoothing the overfitting artifacts of $\xgb$}\label{sec:smoothing_xgb}

Blackbox machine learning algorithms like $\xgb$ often feature artifacts of overfitting as discussed in Section~\ref{sec:overfitting_xgb}.
The flexibility of these algorithms that allows them to capture essential nonlinearities and interactions, which improves predictive performance, also makes them prone to overfitting.
Monotonic constraints can ameliorate this, but we cannot use monotonic constraints for multinomial $\xgb$ because, for instance, the probability that a drive results in a field goal is not monotonic in any covariate.
Thus, in this section, we explore other ways of mitigating these artifacts of overfitting.

\subsection{$\xgb$ regression with monotonic constraints}\label{sec:xgbr}

We can use monotonic constraints to mitigate overfitting if we model $\ep$ using regression rather than classification.
Regression involves directly modeling $\ep$, with outcome variable the net points of the drive $\pts(y)$, rather than modeling drive outcome probabilities as we did previously (see Equation~\eqref{eqn:p_to_ep}).
Because $\ep$ itself is monotonic in certain variables, we can impose monotonic constraints on a regression model.
In particular, we fit a weighted $\xgb$ regression model with monotonic constraints for each pertinent variable (we expect $\ephat$ to be monotonic decreasing in yardline, yards to go, point spread, and defensive team timeouts remaining and monotonic increasing in offensive team timeouts remaining).
We use the same covariates and row weights as in our weighted multinomial $\xgb$ model from Section~\ref{sec:models_dep_structure}.

\begin{figure}[hbt!]
    \centering{}
    \includegraphics[width=0.75\textwidth]{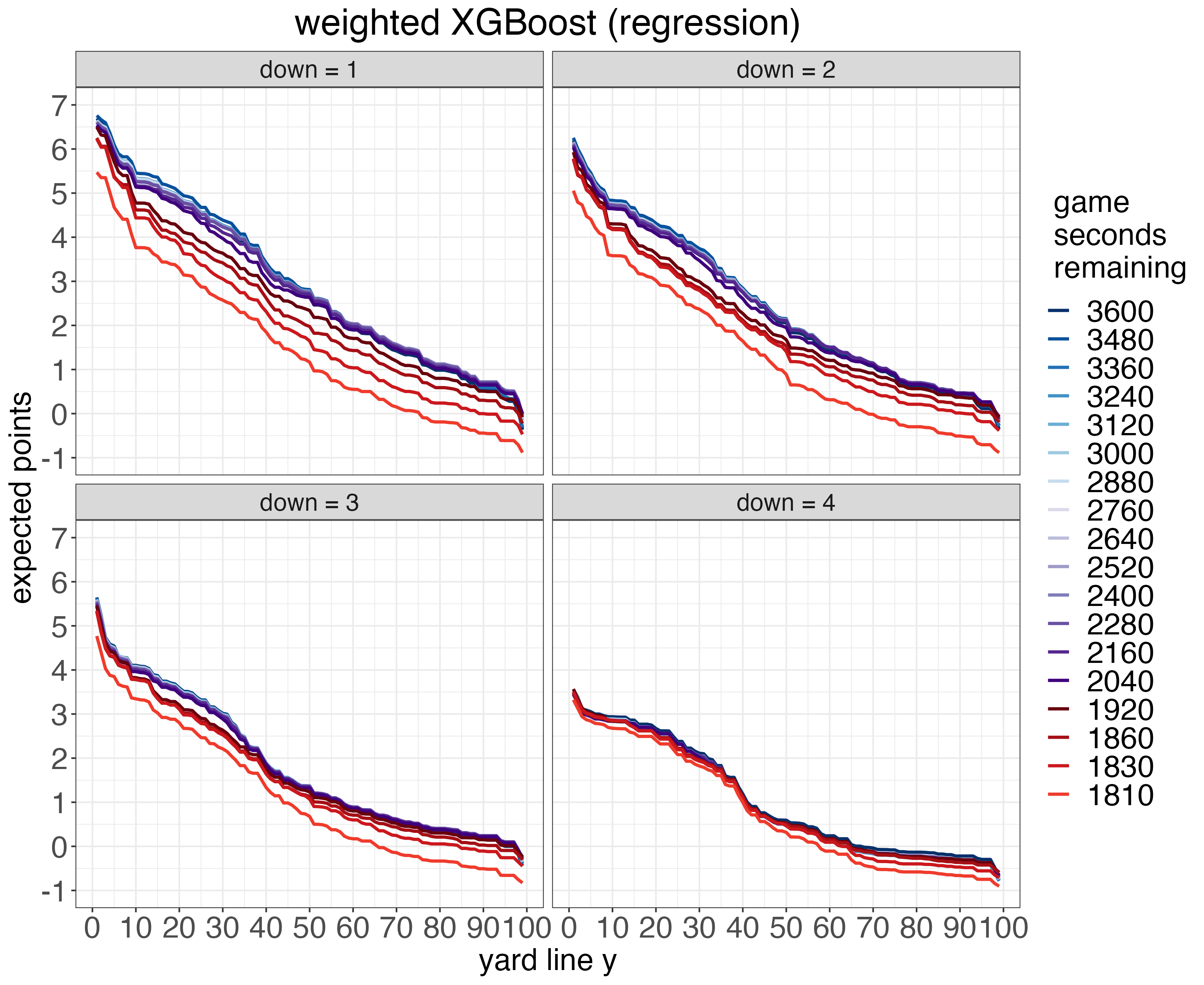}
    \caption{
        $\ephat$ ($y$-axis) according to weighted $\xgb$ regression as a function of yardline ($x$-axis), half seconds remaining (color), and down (facet), holding the other covariates fixed.
    }
    \label{fig:xgbr_time_down}
\end{figure}

Though $\xgb$ regression with monotonic constraints overfits less than multinomial $\xgb$ by construction, it is less accurate (see Table~\ref{table:model_eval_xgbr}).
In other words, modeling the drive outcome probabilities adds predictive power to $\ep$ models.
We suspect this is because it is more difficult to  capture certain nonlinearities and interactions between variables for regression than classification, which we explore in Figure~\ref{fig:xgbr_time_down}.
In the figure, $\xgb$ regression misses that $\ephat$ should be zero at the end of the half in opponent's territory for all downs.
Multinomial $\xgb$ captures this (see Figure~\ref{fig:wxgb_time_down} in Appendix~\ref{app:ep_model_details}) because it directly models the probability that the drive ends in a ``no score'' as a function of game-state.
\begin{table}[htb!]
\centering
\begin{tabular}{lll} \hline
  model $\ephat$ & $\rmse(\ephat, \Dtest)$ \\ \hline 
  weighted multinomial $\xgb$ & $2.593 \pm 0.0017$ \\ 
  weighted $\xgb$ regression & $2.598 \pm 0.0017$  \\ \hline
\end{tabular}
\caption{
$\ep$ model evaluation results for $\xgb$ classification versus regression approaches (plus-minus twice the standard error).
See Formula~\eqref{eqn:test_rmse} for a definition of $\rmse$.
}
\label{table:model_eval_xgbr}
\end{table}


\subsection{Multinomial logistic regression}\label{sec:mlr}

In Section~\ref{sec:xgbr} we found that modeling the drive outcome probabilities, rather than directly modeling the net points of the drive, produces more accurate $\ep$ models.
Thus, we focus on modeling the probability of each potential drive outcome in a way that reduces the overfitting of $\xgb$.
Here, we try a multinomial logistic regression model.
The simple parametric functional form of this model, consisting of a relatively small number of linear terms, won't produce artifacts of overfitting.
The parameters of this model should enforce monotonicity in pertinent variables simply by their signs being correct.
We detail the particular functional form of the best (weighted) multinomial logistic regression model we could find in Appendix~\ref{app:ep_model_details}.

\begin{figure}[hbt!]
    \centering{}
    \subfloat[\centering ]{
        {\includegraphics[width=0.5\textwidth]{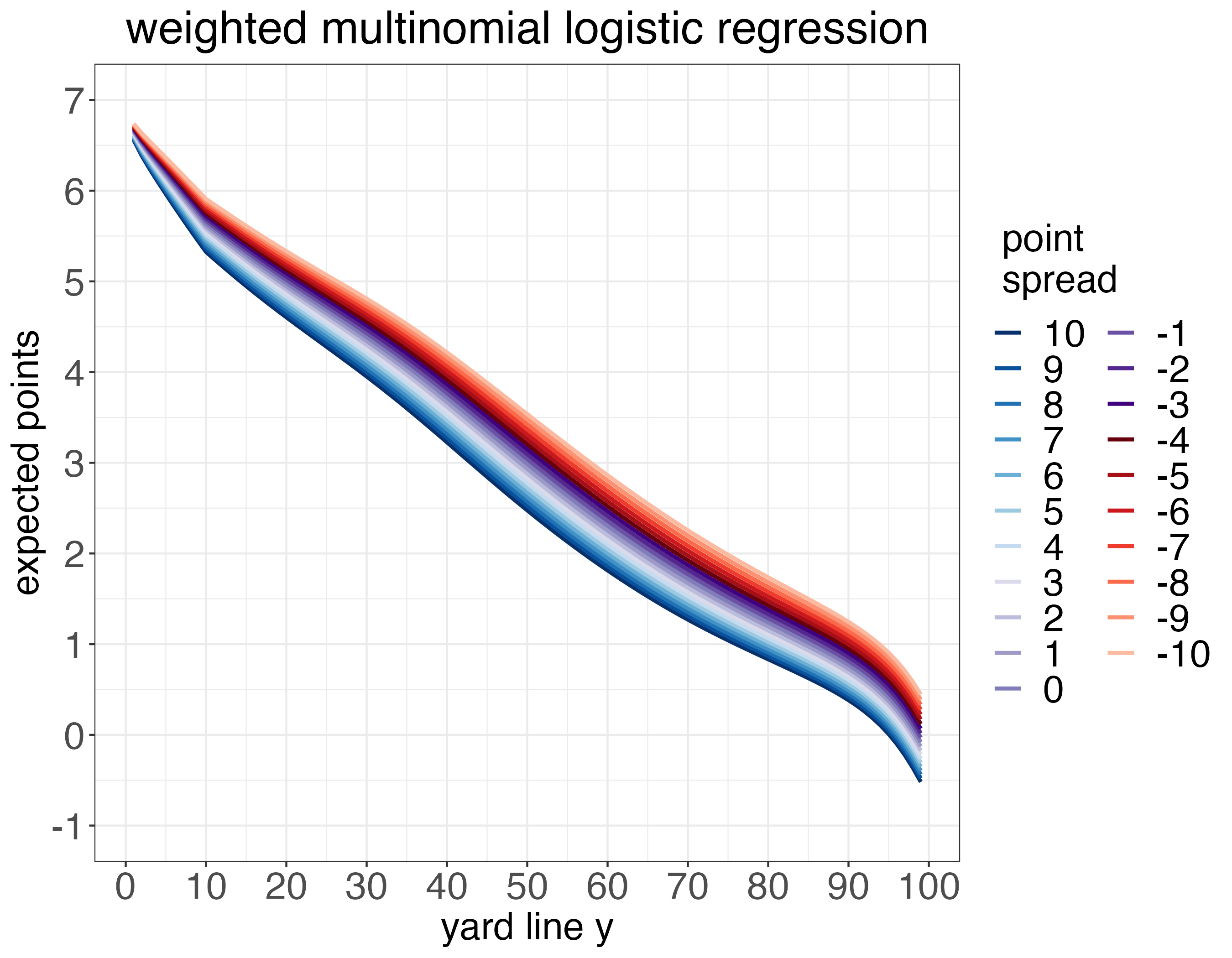}}
        \label{fig:wmlr_pointspread}
    }
    \subfloat[\centering ]{
        {\includegraphics[width=0.5\textwidth]{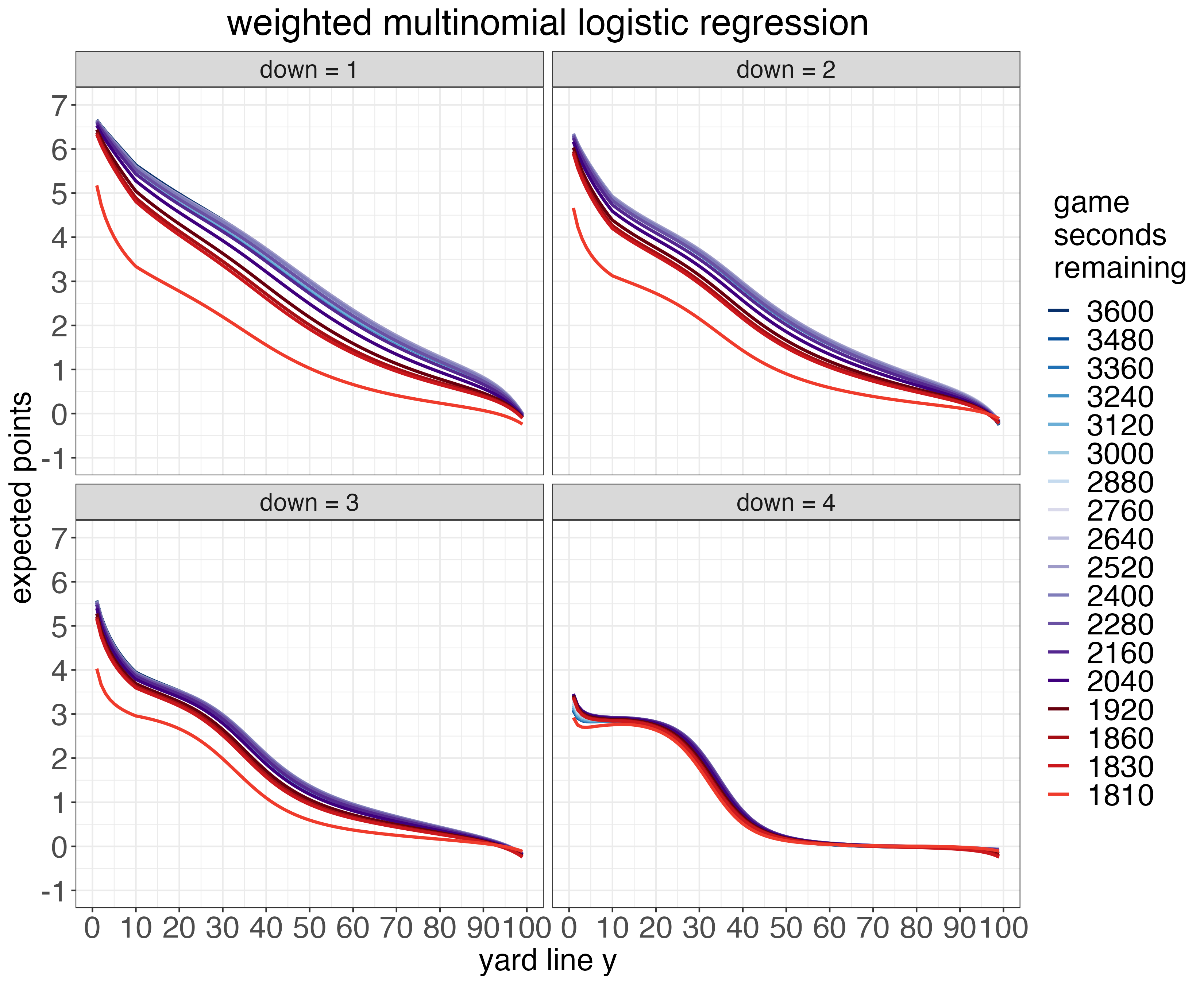}}
        \label{fig:wmlr_time_down}
    }
    \caption{
        Figure (a): $\ephat$ ($y$-axis) according to a weighted multinomial logistic regression model as a function of yardline ($x$-axis) and point spread (color), holding the other covariates fixed.
        Figure (b): $\ephat$ ($y$-axis) according that model as a function of yardline ($x$-axis), half seconds remaining (color), and down (facet), holding the other covariates fixed.
    }
    \label{fig:wmlr_viz}
\end{figure}

In Figure~\ref{fig:wmlr_viz} we visualize this model, which resembles a smoothed version of weighted multinomial $\xgb$ and doesn't exhibt artifacts of overfitting.
But, it is less accurate than multinomial $\xgb$, displayed in Table~\ref{table:model_eval_mlr}, which suggests that there are nonlinearities and interactions that are not captured by the multinomial logistic regression model.

\begin{table}[htb!]
\centering
\begin{tabular}{lll} \hline
  model $\bphat$ & $\logloss(\bphat, \Dtest)$ & $\rmse(\ephat, \Dtest)$ \\ \hline 
  weighted multinomial $\xgb$ & $0.7506 \pm 0.0006$ & $2.593 \pm 0.0017$ \\ 
  weighted multinomial logistic regression & $0.7584 \pm 0.0006$ & $2.601 \pm 0.0017$ \\ \hline
\end{tabular}
\caption{Model evaluation results (plus-minus twice the standard error).
}
\label{table:model_eval_mlr}
\end{table}


We have found a fundamental trade-off between multinomial $\xgb$ and multinomial logistic regression: the former captures essential nonlinear and interacting relationships between variables, yielding accurate but overfit estimates, while the latter is smoother but doesn't capture important relationships in the data, yielding a less accurate model.
The crux of this trade-off is that the representational flexibility of black-box machine learning models, which makes them a popular choice for many applications, also makes them prone to overfitting.
This is a manifestation of the bias-variance trade-off: blackbox machine learning models have lower bias and higher variance than simpler parametric models, which introduce some bias in the form of parametric assumptions in order to reduce variance and smooth the model. 
How to balance this trade-off and smooth the artifacts of overfitting in blackbox machine learning models remains a fascinating open question.

\subsection{Catalytic prior}\label{sec:catalytic}

\begin{figure}[hbt!]
    \centering{}
    \includegraphics[width=0.75\textwidth]{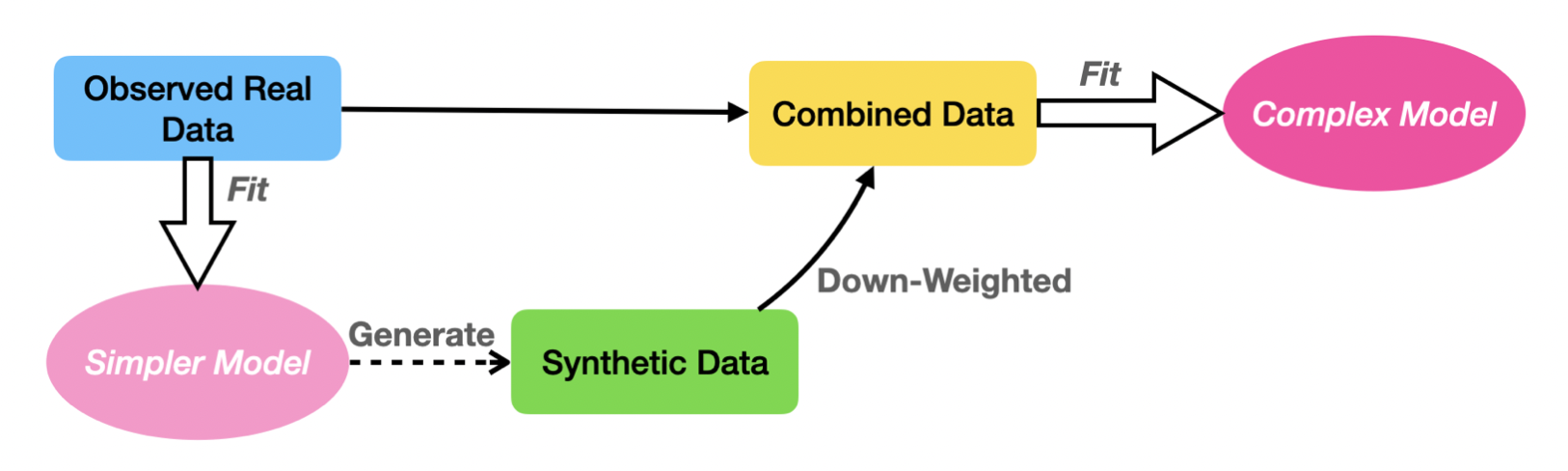}
    \caption{
        The catalytic modeling process \citep{catalytic_prior}.
    }
    \label{fig:catalytic}
\end{figure}

We explore balancing this trade-off and smoothing $\xgb$ models using a catalytic prior, which involves imputation from a simpler model \citep{catalytic_prior_20,catalytic_prior}.
Previously, catalytic priors were introduced in those papers in the context of linear models.
Here, we extend those ideas to our context of machine learning.
The idea is to shrink the complex multinomial $\xgb$ model towards the simpler smoother multinomial logistic regression model.
We accomplish this by augmenting the $\xgb$ training dataset with synthetic data generated from a ``catalytic prior'' model, which in this case is the simpler logistic model.
In Figure~\ref{fig:catalytic} we visualize the catalytic modeling process on a high level \citep{catalytic_prior}.

In Algorithm~\ref{algo:catalyticPrior} we detail the catalytic modeling fitting algorithm in our context of expected points models.
The catalytic modeling process proceeds as follows.
We let the target model be weighted multinomial $\xgb$ and the catalytic prior model be weighted multinomial logistic regression.
We generate (approximately) $M$ synthetic game-states by re-sampling drives with replacement.
Then, for each of these synthetic game-states we impute a synthetic outcome from the catalytic prior model, which we fit from the observed training dataset.
$M$ should be large enough to allow the synthetic dataset to capture nuances of the catalytic prior model, but should also be small enough to keep the model training process relatively quick.
We then let the total weight of the smooth synthetic data be a fraction $\phi$ of the total weight of the observed data.
$\phi$ controls the amount that the target model shrinks to the catalytic prior.
Finally, we fit the target model from a combined dataset consisting of the observed and synthetic data.

\begin{algorithm}[hbt!]
    \caption{Fit a catalytic machine learning model.}  
    \label{algo:catalyticPrior}
    \begin{algorithmic}[1]
        \State \textbf{Input:} 
          \begin{itemize}
            \item $\Dtrain = (\bXtrain, \bytrain, \bwtrain)$, the training dataset  consisting of a feature matrix $\bXtrain$, an outcome vector $\bytrain$, and a vector of row weights $\bwtrain$.
            \item $M$, the (approximate) number of synthetic datapoints.
            \item $\phi$; we let the total weight of the generated synthetic data be the fraction $\phi$ of the total weight of the observed data.
            \item $\D \mapsto \bphatCat(\D)$, the catalytic prior model fitting algorithm (e.g., weighted multinomial logistic regression).
            \item $\D \mapsto \bphatTarget(\D)$, the target model fitting algorithm (e.g., weighted multinomial $\xgb$).
          \end{itemize}

        \State Fit the catalytic prior model from the training dataset, $\bphatCat \leftarrow \bphatCat(\Dtrain)$.
        \State Generate a matrix $\bX_\ast$ consisting of (approximately) $M$ synthetic game-states. 
        To do so, sample drives with replacement from the original game-state training matrix $\bXtrain$ until the number of rows exceeeds $M$.
        \State Generate a vector $\by_\ast$ consisting of (approximately) $M$ synthetic outcomes as follows.
        For each synthetic game-state $\bx_\ast \in \bX_\ast$, generate a synthetic outcome $y_\ast$ from $\bphatCat(\bx_\ast)$.
        \State Let $W = \mathrm{sum}(\bwtrain)$ denote the sum of the row weights of the training dataset. Create a vector $\bw_\ast$ consisting of row weights for the synthetic data such that the weight of each synthetic datapoint is $\phi\cdot W / \mathrm{nrow}(\bX_\ast)$. With this weighting scheme, the total weight of the synthetic data is a fraction $\phi$ of the total weight of the observed data.
        \State Combine the observed training dataset and the synthetic dataset, $\D_{\ast\ast} \leftarrow ([\bXtrain, \bX_\ast], [\bytrain, \by_\ast], [\bwtrain, \bw_\ast])$.
        \State Fit the target model from the combined dataset, $\bphatTarget \leftarrow \bphatTarget(\D_{\ast\ast})$.
        \State \textbf{Return} the fitted target model $\bphatTarget$.
    \end{algorithmic}
\end{algorithm}

\begin{figure}[hbt!]
    \centering{}
    \includegraphics[width=0.75\textwidth]{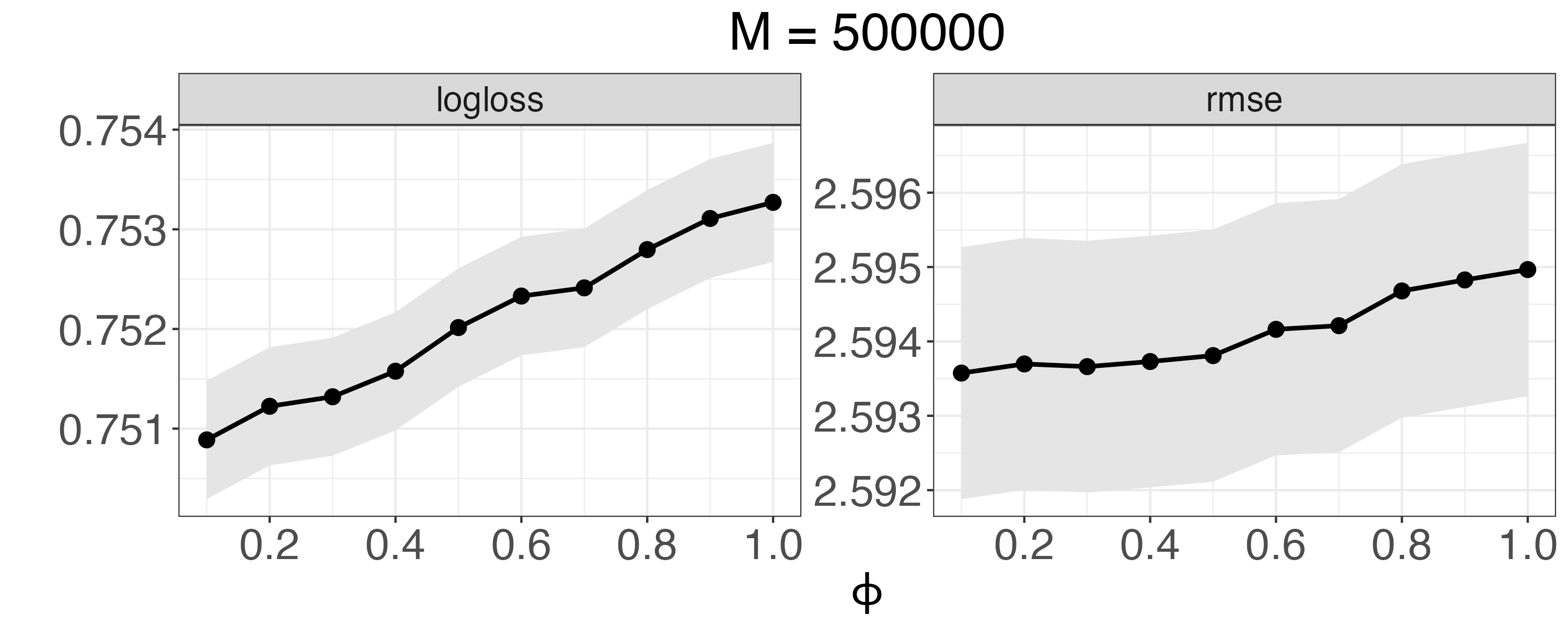}
    \caption{
      Accuracy ($y$-axis) via $\logloss(\bphat, \Dtest)$ (left) and $\rmse(\ephat, \Dtest)$ (right) of our catalytic model as a function of the ratio $\phi$ of weight of synthetic data relative to that of observed data ($x$-axis).
      The gray shaded regions denote plus/minus twice the standard error.
    }
    \label{fig:catalytic_accuracy}
\end{figure}

\begin{figure}[hbt!]
    \centering{}
    \subfloat[\centering ]{
        {\includegraphics[width=0.5\textwidth]{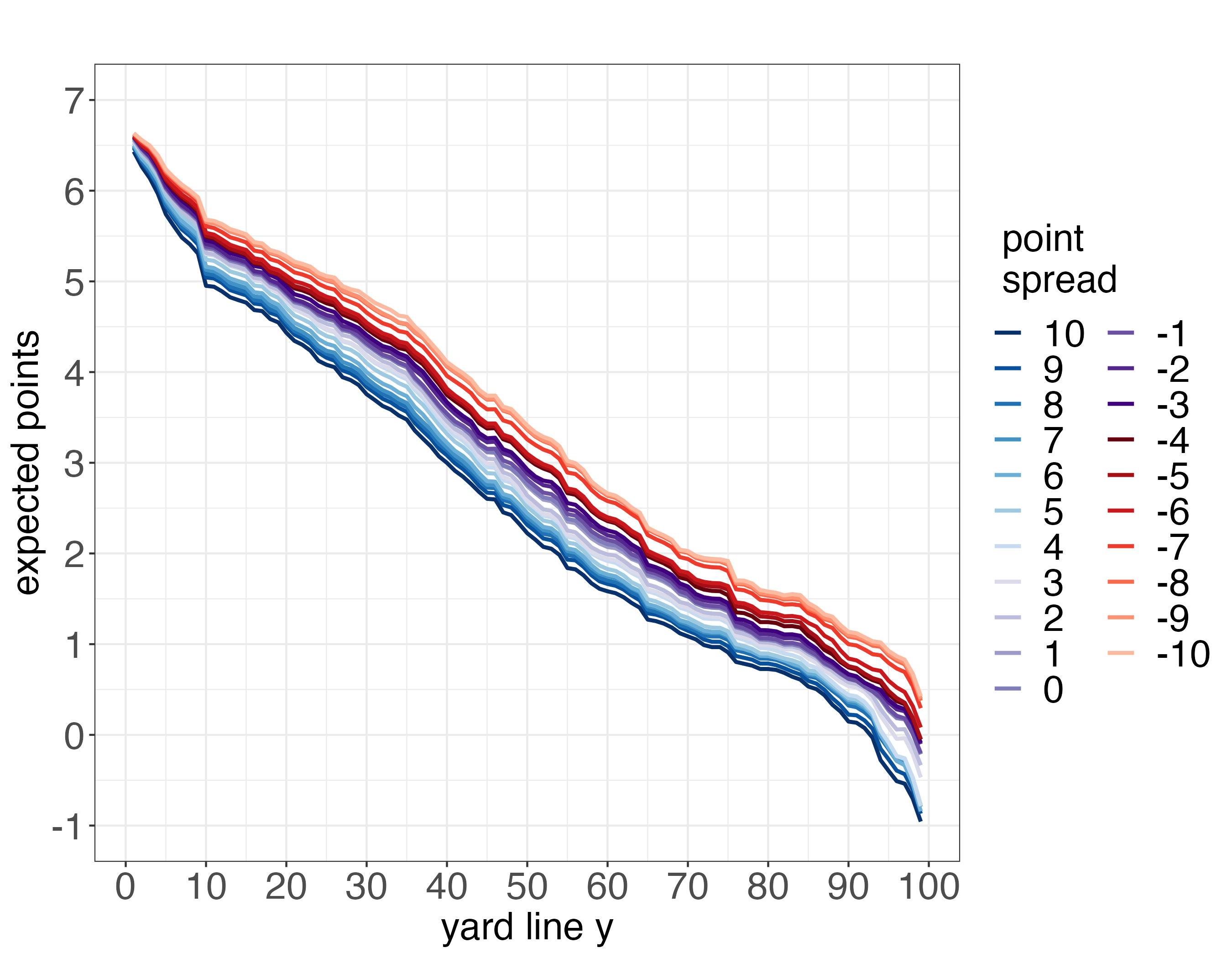}}
        \label{fig:catalytic_tq_L}
    }
    \subfloat[\centering ]{
        {\includegraphics[width=0.5\textwidth]{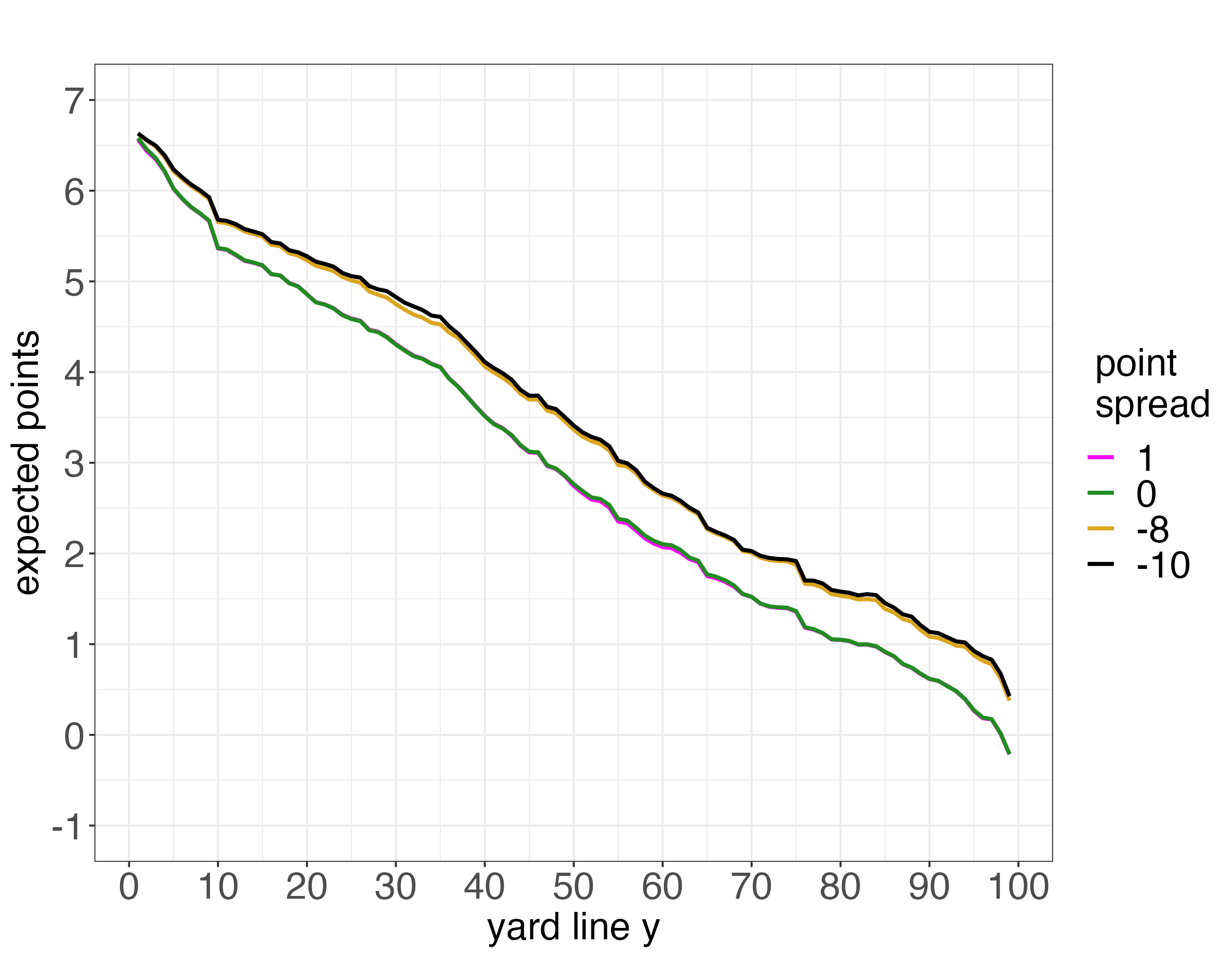}}
        \label{fig:catalytic_tq_R}
    }
    \caption{
        $\ephat$ ($y$-axis) according to our catalytic model with $M=500,000$ and $\phi=1$ as a function of yardline ($x$-axis) and pre-game point spread relative to the team with possession (color), holding the other covariates fixed.  In (b) we see that the artifacts of overfitting from Figure~\ref{fig:wxgb_tq} in Section~\ref{sec:overfitting_xgb} have been smoothed over, as $\ephat$ is now monotonic decreasing in point spread.
    }
    \label{fig:catalytic_tq}
\end{figure}

We use Algorithm~\ref{algo:catalyticPrior} to fit this catalytic model with $M=500,000$ synthetic datapoints for various values of $\phi$.
As the ratio $\phi$ of the weight of smoothed synthetic data relative to that of observed data increases, the catalytic model gets less accurate linearly, as shown in Figure~\ref{fig:catalytic_accuracy}.
But, as $\phi$ increases, the artifacts of overfitting get smoothed over.
The lowest value of $\phi$ for which we no longer find artifacts of overfitting when visualizing the model is about $\phi = 1$, for which the synthetic data and observed data have equal total weight.
We visualize the catalytic model with $\phi=1$ in Figure~\ref{fig:catalytic_tq}. 
Indeed, we see that the artifacts of overfitting described in Figure~\ref{fig:wxgb_tq} in Section~\ref{sec:overfitting_xgb} have been smoothed over; the catalytic model is monotonic decreasing in point spread.
These figures exemplify the trade-off that our catalytic model balances: we can smooth over artifacts of overfitting at the expense of accuracy.

\section{Player evaluation results}\label{sec:player_eval_results}

Football analysts are interested in expected points because it provides a means for player evaluation and strategic decision-making.
Recently, the basis of mathematical strategic decision-making has pivoted to making the decision that maximizes estimated win probability, not estimated expected points \citep{brill2024analytics}.
But, today expected points added ($\epa$) is still a cornerstone of player evaluation.
The success of a play is defined by the change in expected points on that play, and a player is judged by the aggregation of his $\epa$ accross all his plays (e.g., $\epa$ per play).

Now that we have discussed issues with expected points models in Section~\ref{sec:problems_trad_ep} and addressed them in Sections \ref{sec:accounting_dep_struct} and \ref{sec:smoothing_xgb}, we can finally evaluate quarterbacks and teams by estimating their $\epa$ per play.
In doing so, we use drive $\ep$ rather than epoch $\ep$ since each play has a more immediate relationship to the outcome of a drive than to the next scoring event.
We adjust for team quality via pre-game point spread to mitigate selection bias.
We weigh each play in the training dataset by the inverse number of plays in its drive to account for the dependency structure of observational football data, which increases model accuracy.
We use a cluster bootstrap to create confidence intervals for a player or team's $\epa$ in order to fully capture uncertainty in the outcome of a drive. 

\begin{figure}[hbt!]
    \centering{}
    \subfloat[\centering ]{
        {\includegraphics[width=0.45\textwidth]{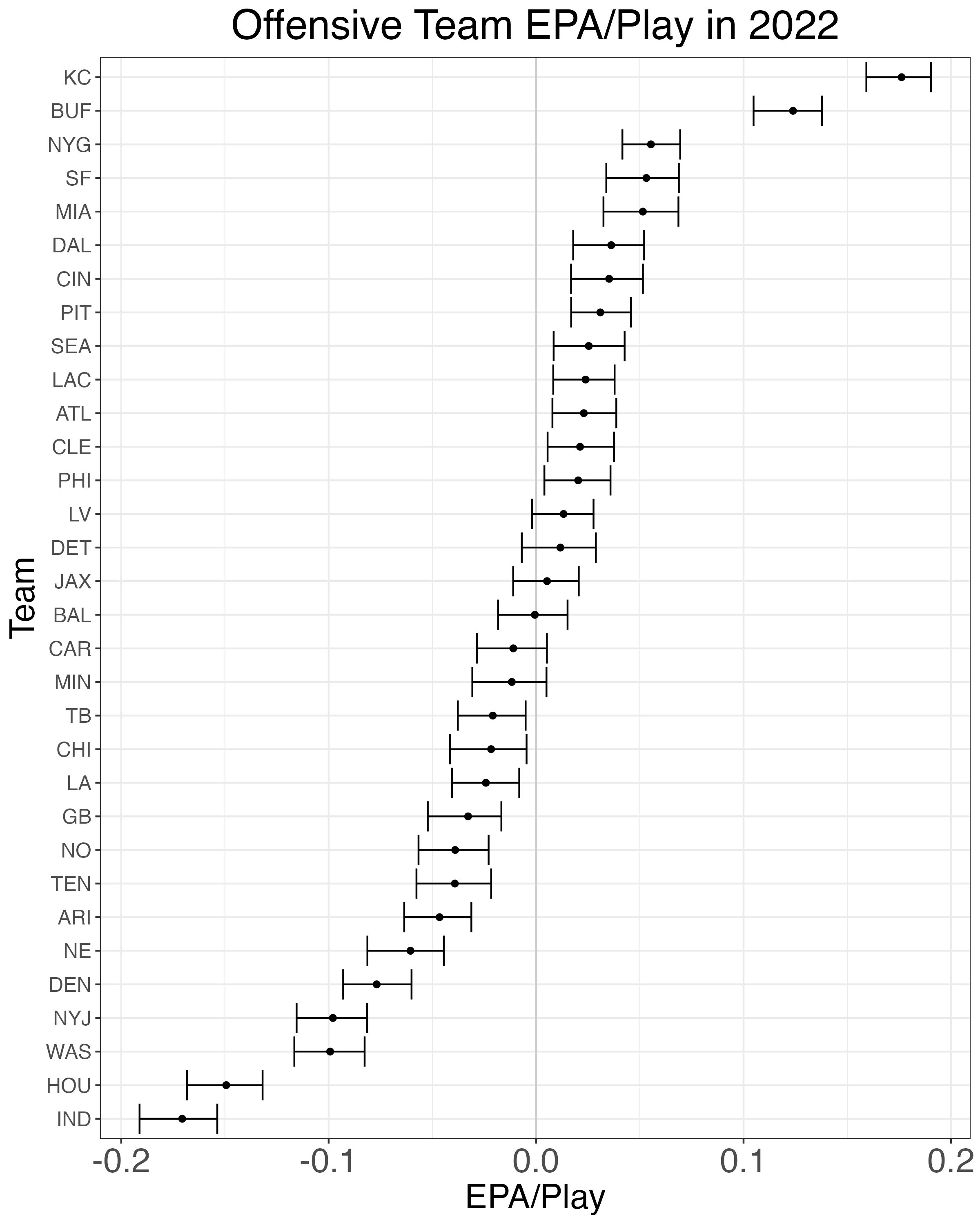}}
        \label{fig:plot_epa_per_play_TM}
    }
    \qquad
    \subfloat[\centering ]{
        {\includegraphics[width=0.45\textwidth]{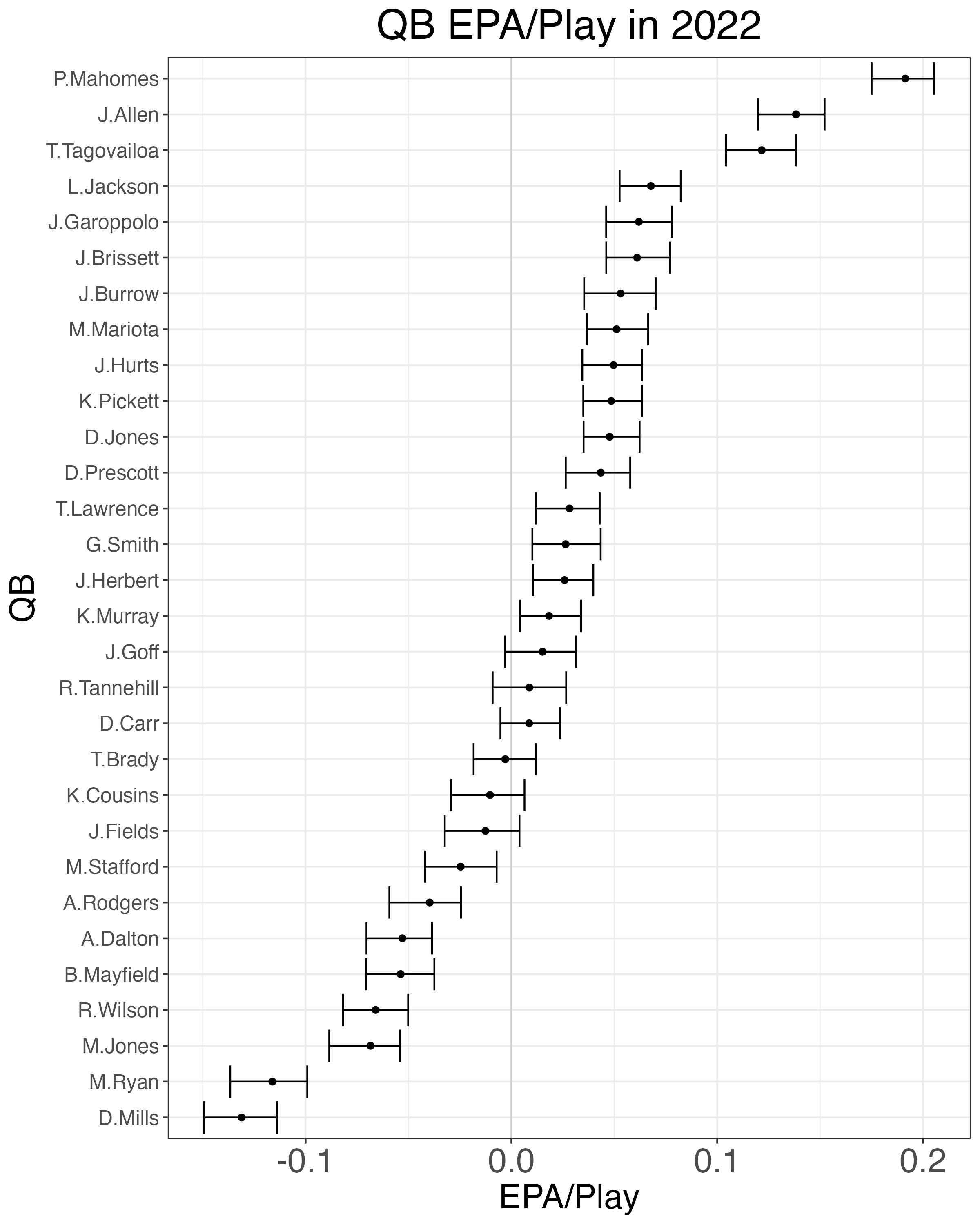}}
        \label{fig:plot_epa_per_play_QB}
    }
    \caption{
        $\epa$ per play for offensive teams (a) and quarterbacks who had at least $500$ pass or run plays (b) in 2022.
        The dots indicate point estimates and the bars indicate $95\%$ bootstrapped confidence intervals.
    }
    \label{fig:epa_per_play_results}
\end{figure}

In Figure~\ref{fig:epa_per_play_results} we visualize $\epa$ per play for offensive teams (Figure~\ref{fig:plot_epa_per_play_TM}) and quarterbacks (Figure~\ref{fig:plot_epa_per_play_QB}) who had at least $500$ pass or run plays.
We set the point spread to be $0$ in generating these $\epa$ estimates so that $\epa$ represents expected points above average.
In 2022 Kansas City had a significantly more efficient offense than Buffalo, who had a significantly more efficient offense than every other team.
Indianapolis and Houston had significantly less efficient offenses than every other team.
Aside from these four outliers, many of the other teams' $\epa$ per play confidence intervals overlap.
Dallas ($6^{th}$) wasn't significantly more efficient than Detroit ($15^{th}$) and Carolina ($18^{th}$) wasn't significantly more efficient than Tennessee ($25^{th}$).
Quarterbacks tell a similar story: in 2022 Mahomes was significantly more efficient than Allen and Tua, who were significantly more efficient than every other quarterback.
Matt Ryan and Davis Mills were significantly less efficient than every other quarterback.
Aside from these five outliers, many of the other quarterbacks' $\epa$ per play confidence intervals overlap.
Lamar Jackson ($4^{th}$) wasn't significantly more efficient than Dak Prescott ($12^{th}$) and Trevor Lawrence ($13^{th}$) wasn't significantly more efficient than Tom Brady ($20^{th}$).
These figures illustrate the need for uncertainty quantification: we need to know whether differences in player or team performance were plausibly due to random chance.

\section{Discussion}\label{sec:discussion}

The most widely used value function across sports analytics is expected points.
Expected points is central to player evaluation and in-game strategic decision-making in American football analytics.
$\ep$ is traditionally framed as a regression problem estimated at the level of plays, which allows $\ep$ models to borrow strength across similar game-states.
The $\ep$ estimates commonly used today arise from blackbox machine learning tools like $\xgb$, which capture essential nonlinearities and interactions between variables.
Despite these strengths of traditional models, in this study we explore some of their weaknesses. 
Notably, they do not consider the dependency structure of observational football, whose outcomes are clustered into drives or epochs.
Prediction sets formed from these models don't achieve adequate marginal coverage.
By not adjusting for team quality, they suffer from selection bias.
Finally, machine learning estimates display counter-intuitive artifacts of overfitting.

We devised expected points models that account for the dependency structure, which improved accuracy, albeit slightly (but statistically significantly).
In particular, we recommend weighing each play in the training dataset by the inverse number of plays in its drive.
To fully account for uncertainty in the outcome of a drive, we used the bootstrap, which produces prediction sets that achieve adequate marginal coverage.
The bootstrap also produces confidence intervals on quantities like $\epa$ per play, allowing us to deduce how granularly to trust differences in $\epa$ per play between players.
In particular, we recommend using a cluster boostrap (sample clusters with replacement).
To mitigate selection bias, we adjusted for team quality using pre-game point spread.
A more elaborate model could include for more refined measures of offensive and defensive quality; perhaps context-neutral $\epa$ per play should also adjust for defensive quality.
Finally, we introduced a novel technique to mitigate overfitting in blackbox machine learning models.
We explored using a catalytic prior to shrink estimates from a machine learning model towards a simpler, smoother, interpretable model by imputing synthetic data.
We found that this approach can smooth over artifacts of overfitting, albeit at the expense of accuracy.

Continuing on the work done in this study, we see two promising avenues for future work.
First, we look forward to a continuted exploration of the efficacy of catalytic priors or other imputation techniques in smoothing or shrinking machine learning models.
In this study, we just provided an example illustrating the promise of catalytic priors, but there is more work to be done to better understand when to use such an approach and how effective it can really be.
Additionally, we look forward to further research into understanding the effect of the dependence structure detailed in this paper on the bias and variance of estimators in other sports analytics examples (e.g., \citet{brill2024exploringdifficultyestimatingwin}).
The clustered nature of observational football data is not unique to expected points models, as similar dependence structures are common across sports analytics.
A similar correlation structure appears in any dataset in which the outcome variable is the final result of some unit of time (i.e., game, quarter, or play) and the observations consist of frames, plays, or actions leading to that end result.
For instance, grading wide receivers using NFL tracking data involves mapping the actions of wide receives across all the frames of a play leading up to the resulting play outcome.

\if0\blind
{
  \section*{Acknowledgments}

The authors acknowledge the High Performance Computing Center (HPCC) at The Wharton School, University of Pennsylvania for providing computational resources that have contributed to the research results reported within this paper. 
} \fi

\bibliography{refs}

\clearpage
\newpage
\begin{center}
{\large\bf SUPPLEMENTARY MATERIAL}
\end{center}
\appendix

\section{Model specification details}\label{app:ep_model_details}

\begin{figure}[hbt!]
    \centering{}
    \includegraphics[width=0.75\textwidth]{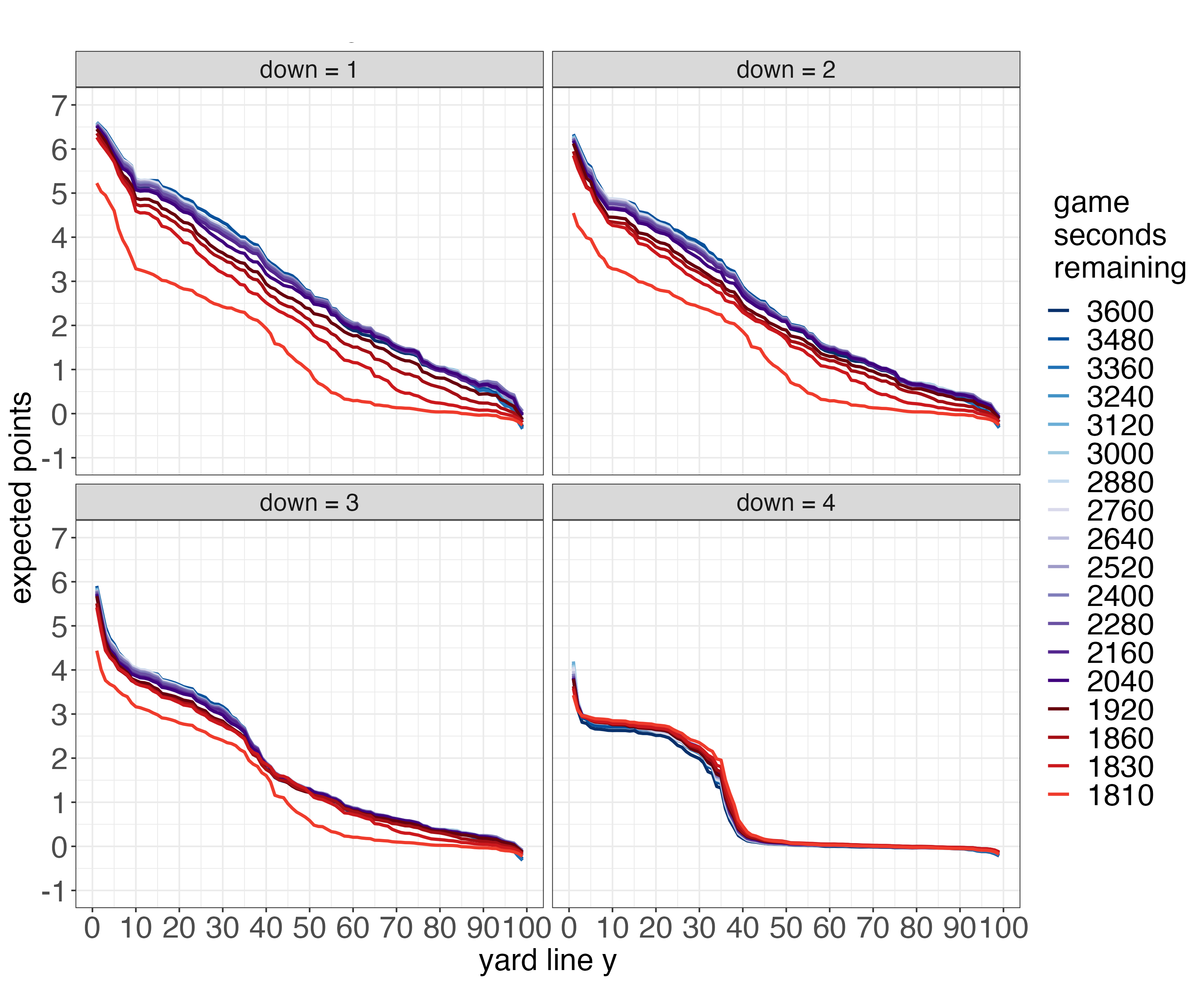}
    \caption{
        $\ephat$ ($y$-axis) according to weighted multinomial $\xgb$ as a function of yardline ($x$-axis), half seconds remaining (color), and down (facet), holding the other covariates fixed.
    }
    \label{fig:wxgb_time_down}
\end{figure}

The $\R$ code for the weighted multinomial logistic regression model we fit in Section~\ref{sec:mlr} and use as a catalytic prior in Section~\ref{sec:catalytic} is
{\footnotesize
\begin{verbatim}
nnet::multinom(
  outcome_drive ~ 
    factor(down):(
     bs(yardline_100, df=5) + bs(half_seconds_remaining, knots=c(30))
    ) +
    log(ydstogo) +
    utm:as.numeric(posteam_timeouts_remaining==0) +
    factor(era) +
    posteam_spread + posteam_spread:yardline_100 +
    I((score_differential <= -11)) + 
    I((score_differential >= 11)) + 
    I((score_differential <= -4)*(game_seconds_remaining <= 900)) +   
    I((-3 <= score_differential & score_differential <= 0)*(game_seconds_remaining <= 900)) + 
    I((1 <= score_differential & score_differential <= 3)*(game_seconds_remaining <= 900)) + 
    I((4 <= score_differential & score_differential <= 10)*(game_seconds_remaining <= 900)),
  weights = drive_weights
).
\end{verbatim}
}

\end{document}